\documentclass[12pt]{article}

\usepackage[superscript]{cite}
\usepackage{comment}
\usepackage{xcolor}
\usepackage{empheq}
\usepackage{amsmath}
\usepackage{amssymb}
\usepackage{ulem,xpatch}
\usepackage{titling}
\usepackage{geometry}

\newcommand{\hide}[1]{\relax}

\newcommand{\Og}{\ensuremath{\Omega}}
\newcommand{\Om}{\ensuremath{\Omega_\mathrm{m}}}
\newcommand{\Qm}{\ensuremath{Q}}

\newcommand{\Oeff}{\ensuremath{\Omega_\mathrm{eff}}}
\newcommand{\Gm}{\ensuremath{\Gamma_\mathrm{m}}}
\newcommand{\Geff}{\ensuremath{\Gamma_\mathrm{eff}}}
\newcommand{\Gopt}{\ensuremath{\Gamma_\mathrm{opt}}}

\newcommand{\oC}{\ensuremath{\omega_\mathrm{cav}}}

\newcommand{\meff}{m}

\newcommand{\chieff}{\chi_\text{eff}}
\newcommand{\chim}{\chi_\text{m}}

\newcommand{\hfb}{h_\text{fb}}
\newcommand{\Tm}{T}

\newcommand{\SFFtot}{\ensuremath{\bar S_{FF}^{\mathrm{tot}}}}
\newcommand{\SFFth}{\ensuremath{\bar S_{FF}^{\mathrm{th}}}}
\newcommand{\SFFaux}{\ensuremath{\bar S_{FF}^{\mathrm{aux}}}}
\newcommand{\SFFqba}{\ensuremath{\bar S_{FF}^{\mathrm{qba}}}}

\newcommand{\Sxximp}{\ensuremath{\bar S_{xx}^{\mathrm{imp}}}}

\newcommand{\Cx}{C_{XX}}
\newcommand{\Cy}{C_{YY}}
\newcommand{\vcrP}{\ensuremath{g_0}}
\newcommand{\nI}{\bar{n}_\mathrm{i}}
\newcommand{\nF}{\bar{n}_\mathrm{f}}
\newcommand{\Xt}{\mathrm{X(t)}}

\newcommand{\Yt}{\mathrm{Y(t)}}

\newcommand{\RftSqAv}{\mathrm{\langle R^2(t)\rangle}}
\newcommand{\XftSqAv}{\mathrm{\langle X^2(t)\rangle}}
\newcommand{\YftSqAv}{\mathrm{\langle Y^2(t)\rangle}}
\newcommand{\RsnftSqAv}{\mathrm{\langle R_{sn}^2(t)\rangle}}
\newcommand{\finRftSqAv}{\mathrm{\langle \widetilde{R}^2(t)\rangle}}

\newcommand{\bn}{\ensuremath{\bar n}}
\newcommand{\bnth}{\ensuremath{\bar n_\mathrm{th}}}
\newcommand{\bnmin}{\ensuremath{\bar n_\mathrm{min}}}
\newcommand{\bncav}{\ensuremath{\bar n_\mathrm{cav}}}
\newcommand{\bnest}{\ensuremath{\bar n_\mathrm{est}}}
\newcommand{\bncavpr}{\ensuremath{\bar n_\mathrm{cav}}}
\newcommand{\bncavaux}{\ensuremath{\bar n_\mathrm{cav}^\mathrm{aux}}}

\newcommand{\mhat}{}

\newcommand{\kB}{\ensuremath{k_\mathrm{B}}}

\newcommand{\ha}{\ensuremath{\mhat a}}
\newcommand{\had}{\ensuremath{\mhat a^\dagger}}

\newcommand{\dhx}{\ensuremath{\delta \mhat x}}

\newcommand{\dhp}{\ensuremath{\delta \mhat p}}

\newcommand{\Gmeas}{\ensuremath{\Gamma_\mathrm{meas}}}
\newcommand{\Gqba}{\ensuremath{\Gamma_\mathrm{qba}}}
\newcommand{\gtot}{\ensuremath{\gamma_\mathrm{tot}}}

\newcommand{\Cq}{\ensuremath{C_\mathrm{q}}}

\newcommand{\Daux}{\ensuremath{\Delta_\mathrm{aux}}}
\newcommand{\kaux}{\ensuremath{\kappa_\mathrm{aux}}}
\newcommand{\Paux}{\ensuremath{P_\mathrm{aux}}}
\newcommand{\vcrAux}{{g_\mathrm{0}}_\mathrm{aux}}

\newcommand{\hx}{\ensuremath{\mhat x}}

\newcommand{\hb}{\ensuremath{\mhat b}}
\newcommand{\hbd}{\ensuremath{\mhat b^\dagger}}

\newcommand{\xzpf}{\ensuremath{x_\mathrm{zpf}}}
\newcommand{\pzpf}{\ensuremath{p_\mathrm{zpf}}}

\newcommand{\vcr}{\ensuremath{g_0}}

\newcommand{\etac}{\ensuremath{\eta_\mathrm{c}}}
\newcommand{\etadet}{\ensuremath{\eta_\mathrm{det}}}

\newcommand{\kpr}{\ensuremath{\kappa}}
\newcommand{\Dpr}{\ensuremath{\Delta}}

\newcommand{\nimp}{\ensuremath{n_\mathrm{imp}}}

\newcommand{\Fth}{\ensuremath{F_\mathrm{th}}}
\newcommand{\Fba}{\ensuremath{F_{\mathrm{ba}}}}
\newcommand{\Ffb}{\ensuremath{F_{\mathrm{fb}}}}
\newcommand{\ximp}{\ensuremath{x_\mathrm{imp}}}
\newcommand{\Ftot}{\ensuremath{F_{\mathrm{tot}}}}
\newcommand{\Sx}{\ensuremath{\bar{S}_{xx}(\Omega)}}
\newcommand{\Sp}{\ensuremath{\bar{S}_{pp}(\Omega)}}
\newcommand{\Sy}{\ensuremath{\bar{S}_{yy}(\Omega)}}

\newcommand{\Simp}{\ensuremath{\bar{S}_{xx}^\mathrm{imp}(\Omega)}}
\newcommand{\Stot}{\ensuremath{\bar{S}_{FF}^\mathrm{tot}(\Omega)}}

\newcommand{\gfb}{\ensuremath{g_\mathrm{fb}}}
\newcommand{\Gammafb}{\ensuremath{\Gamma_\mathrm{fb}}}
\newcommand{\Ogfb}{\ensuremath{\Omega_\mathrm{fb}}}
\newcommand{\hfbaux}{\ensuremath{h_\mathrm{aux}}}
\newcommand{\hfbmain}{\ensuremath{h_\mathrm{main}}}

\newcommand{\eref}[1]{(\ref{#1})}
\newcommand{\nocontentsline}[3]{}
\newcommand{\tocless}[2]{\bgroup\let\addcontentsline=\nocontentsline#1{#2}\egroup}

\newenvironment{sciabstract}{%
\begin{quote} \bf}
{\end{quote}}

\title{\bf \vspace{-2cm} Measurement-based quantum control \\of mechanical motion} 
\author{\normalsize{Massimiliano Rossi$^{1, 2,\ast}$, David Mason$^{1,2,\ast}$, Junxin Chen$^{1,2,\ast}$, Yeghishe Tsaturyan$^{1}$ \& Albert Schliesser$^{1,2,\dagger}$}\\
\vspace{7mm}\small{\it $^{1}$Niels Bohr Institute, University of Copenhagen, 2100 Copenhagen, Denmark}\\
\small{\it$^{2}$Center for Hybrid Quantum Networks (Hy-Q), Niels Bohr Institute,}\\
\small{\it University of Copenhagen, 2100 Copenhagen, Denmark}\\
\vspace{7mm}\small{$^\ast$these authors contributed equally to this work}\\
\small{$^\dagger$to whom correspondence should be addressed; e-mail:  albert.schliesser@nbi.ku.dk}}
\date{}

\pretitle{\begin{flushleft}\LARGE} 
\posttitle{\end{flushleft}}
\preauthor{\begin{flushleft}\large} 
\postauthor{\end{flushleft}}

\begin{document}

\maketitle

\begin{sciabstract}

Controlling a quantum system based on the observation of its dynamics is inevitably complicated by the backaction of the measurement process.
Efficient measurements, however, maximize the amount of information gained per disturbance incurred.
Real-time feedback then enables both canceling the measurement's backaction and controlling the evolution of the quantum state.
While such measurement-based quantum control has been demonstrated in the clean settings of cavity and circuit quantum electrodynamics, 
its application to motional degrees of freedom has remained elusive.
Here we show measurement-based quantum control of the motion of a millimetre-sized membrane resonator. 
An optomechanical transducer resolves the zero-point motion of the soft-clamped 
resonator in a fraction of its millisecond coherence time, 
with an overall measurement efficiency close to unity.
We use this position record to feedback-cool a resonator mode to its quantum ground state (residual thermal occupation $\bn = 0.29\pm 0.03$), 9~dB below the quantum backaction limit of sideband cooling, and six orders of magnitude below the equilibrium occupation of its thermal environment.
This realizes a long-standing goal in the field, and adds position and momentum to the degrees of freedom amenable to measurement-based quantum control, with potential applications in quantum information processing and gravitational wave detectors.

\end{sciabstract}

\newpage
Controlling the state of a quantum system is a delicate task, since observation of the system will inevitably perturb it\cite{Braginsky:1992aa,Clerk:2010ab}.
\textit{Coherent} quantum control avoids this issue, by coupling the system to another ``controller'' quantum system in such a way that the joint system converges to the target state without the need for measurement---at the expense of quantum resources in the controller.
\textit{Measurement-based} quantum control\cite{Jacobs:2014aa,Wiseman:2009aa,Zhang:2017aa} is based on a different paradigm.
It exerts control by measuring the quantum state, and  applying feedback that depends on the measurement outcome, much alike classical control systems.
In the quantum regime, however, the effect of the measurement's backaction must be taken into account, and effectively canceled.
This requires an overall measurement efficiency $\eta$---in essence the amount of information gained per decoherence induced---close to unity, a challenging demand yet met only with the impeccable systems of cavity and circuit QED\cite{Sayrin:2011aa,Vijay:2012aa} (e.g. $\eta=40\,\%$ in ref.\ \cite{Vijay:2012aa}).

To prepare high-purity \textit{motional} quantum states, researchers have traditionally relied on sideband cooling, a form of coherent quantum control.
An engineered quantum optical bath acts as controller, to which the motional degree of freedom couples through optical forces.
The motion thermalizes to this bath, at a temperature determined by the forces' quantum fluctuations. This temperature sets a fundamental limit to sideband cooling.  
In optomechanics, this limit requires that the cavity linewidth resolves the motional sidebands to enable ground state cooling with coherent light\cite{Aspelmeyer:2014aa}.
Systems operating in this regime have been prepared close to the ground state\cite{Chan:2011aa,Teufel:2011aa}, and a recent work demonstrated cooling 2-dB below the sideband cooling limit by squeezing the electromagnetic vacuum fluctuations\cite{Clark:2017aa}.

Within the paradigm of measurement-based quantum control, feedback cooling\cite{Mancini:1998aa,Cohadon:1999aa} can overcome this limit, given a sufficiently efficient measurement. 
Several communities, including atomic physics, optomechanics, and gravitational wave astrononomy, have therefore explored this protocol.
Yet in spite of a two-decade effort with mechanical systems as diverse as 
trapped atoms\cite{Kubanek:2009aa},
ions\cite{Bushev:2006aa}, 
micro- and nanoparticles\cite{Li:2011aa,Jain:2016aa},
cantilevers\cite{Poggio:2007aa,Kleckner:2006aa},
nanomechanical resonators\cite{Wilson:2015aa,Lee:2010aa,LaHaye:2004aa,Gavartin:2012aa}, 
mirror modes\cite{Cohadon:1999aa}, 
and gravitational wave detector masses\cite{Vinante:2008aa,Abbott:2009aa}
the goal of ground-state cooling, an elementary form of quantum control, has never been reached.

This is, chiefly, because the measurements were too weak ($\Gmeas \ll \gamma$) and/or the detection too inefficient ($\Gmeas \ll \Gqba$), to realise an overall measurement efficiency 
\begin{equation}
  \eta = \frac{\Gmeas}{\Gqba+\gamma} \sim 1.
\end{equation}
Here, $\Gmeas$ is the measurement rate\cite{Clerk:2010ab}, and $\Gqba$ and $\gamma$ are the motional decoherence rates due to the measurement quantum backaction and coupling to the environment, respectively. 
The closest approach, to our knowledge, has been reported by Wilson et al.\cite{Wilson:2015aa}, performing a feedback experiment on a nanomechanical resonator with $\eta=0{.}9\%$.
In contrast to these previous attempts, we perform motion measurement that is sufficiently strong and efficient, to reach up to $\eta=56\%$.
This is enabled by an extremely precise displacement measurement, which realises the yet closest approach (within $35\%$) to the Heisenberg measurement-disturbance uncertainty limit and the standard quantum limit.

\paragraph {Experimental setting}

Concretely, we study the drumhead-like motion of a highly tensioned, millimetre-sized but 20-nm-thin, $\mathrm{Si_3N_4}$ membrane (Fig.~1).
The resonance mode of interest is confined to a defect within a phononic crystal (PnC), created by patterning a periodic array of holes into the membrane.
The frequency ${\Om/2\pi=1.14~\mathrm{MHz}}$ of the defect mode lies in the bandgap of the surrounding PnC, minimizing radiative leakage of mechanical energy into the surrounding structure. 
The gentle confinement by the PnC simultaneously reduces mode curvature compared to membranes clamped to a rigid substrate.
As we have recently shown\cite{Tsaturyan:2017aaa}, such ``soft clamping'' dramatically suppresses mechanical energy dissipation ($\Gm$) and enables ultrahigh quality factors $Q=\Om/\Gm$:
indeed, we find $Q=1.03\times10^9$ in ringdown measurements, carefully ruling out artefacts (see Supplementary).
This corresponds to a mechanical coherence time $\gamma^{-1}\approx (\bnth \Gm)^{-1}=\hbar Q/(k_\mathrm{B} T)$ on the order of $1\,\mathrm{ms}$ ($\hbar$ reduced Planck's constant, $k_\mathrm{B}$ Boltzmann constant, $T$ environment temperature, $\bnth$ thermal bath occupation) even at the moderate cryogenic temperatures ($T\sim10\,\mathrm{K}$, $\bnth\sim\mathcal{O}(10^5)$) at which all reported experiments are conducted. 

The membrane is introduced in a 1.6-mm-long, high-finesse Fabry-P\'erot resonator, so that displacement by its zero-point-amplitude $\xzpf=\sqrt{\hbar/2m\Om}$ ($m$ resonator mass) dispersively shifts\cite{Thompson:2008aaa,Nielsen:2017aaa} (see Supplementary) the optical mode frequency by $\vcr$, the vacuum optomechanical coupling rate.
Populating the cavity with a coherent field of average photon number $\bncav$ then leads to the field-enhanced coupling $g=\vcr\sqrt{\bncav}$ in a linearized, QND-type interaction Hamiltonian $H'=-\hbar g(\had+\ha)(\hbd+\hb)$ between the shifted annihilation (creation) operators $\ha$ ($\had$) and $\hb$ ($\hbd$)  of cavity field and mechanical motion, respectively\cite{Aspelmeyer:2014aa,Bowen:2016aa}.
A probe laser (red, Fig.~1a) is used to probe the frequency fluctuations of an optical cavity mode of linewidth $\kappa/2\pi=15.9$ MHz.
We measure mechanical position by monitoring the phase of the transmitted light, using balanced homodyne detection.  
In an unresolved sideband system ($\kappa \gg \Om$), this measurement occurs at a rate\cite{Bowen:2016aa} $\Gmeas=4\etadet g^2/\kappa$, for a detection efficiency $\etadet$.
Careful optimization of the entire detection chain (see Supplementary) leaves us with $\etadet=$~77\%.

In addition, we frequently use an auxiliary laser (blue, Fig.~1a) which populates a different longitudinal cavity mode (linewidth $\kaux/2\pi=$~12.9~MHz) and has a polarization orthogonal to the probe beam, to avoid unwanted interference. It has several purposes, including laser cooling and, in combination with an amplitude modulator, exerting a force on the mechanical resonator via radiation pressure. Its exact role is specified in each section, describing the different experiments performed.

To gauge the possible strength of the measurement, we perform  optomechanically induced transparency (OMIT) measurements\cite{Weis:2010aaa}  to extract the field-enhanced optomechanical coupling $g$.
We find (Fig.~1e) values up to $g/2\pi=$~329~kHz, which suggests that the effect of measurement-induced quantum backaction ($\Gqba= 4g^2/\kappa$) exceeds the thermal decoherence rate ($\gamma$) by a large margin.
This is captured by the quantum cooperativity parameter\cite{Aspelmeyer:2014aa,Bowen:2016aa} {$\Cq=\Gqba/\gamma$, reaching up to $\Cq=119$.} %
{We} can therefore expect a close-to-unity overall measurement efficiency $\eta=\etadet/ (1+1/\Cq)$, as required for successful quantum control.

\begin{figure}
\renewcommand{\figurename}{{\bf Fig.}}
\begin{center}
\includegraphics[scale=1.0]{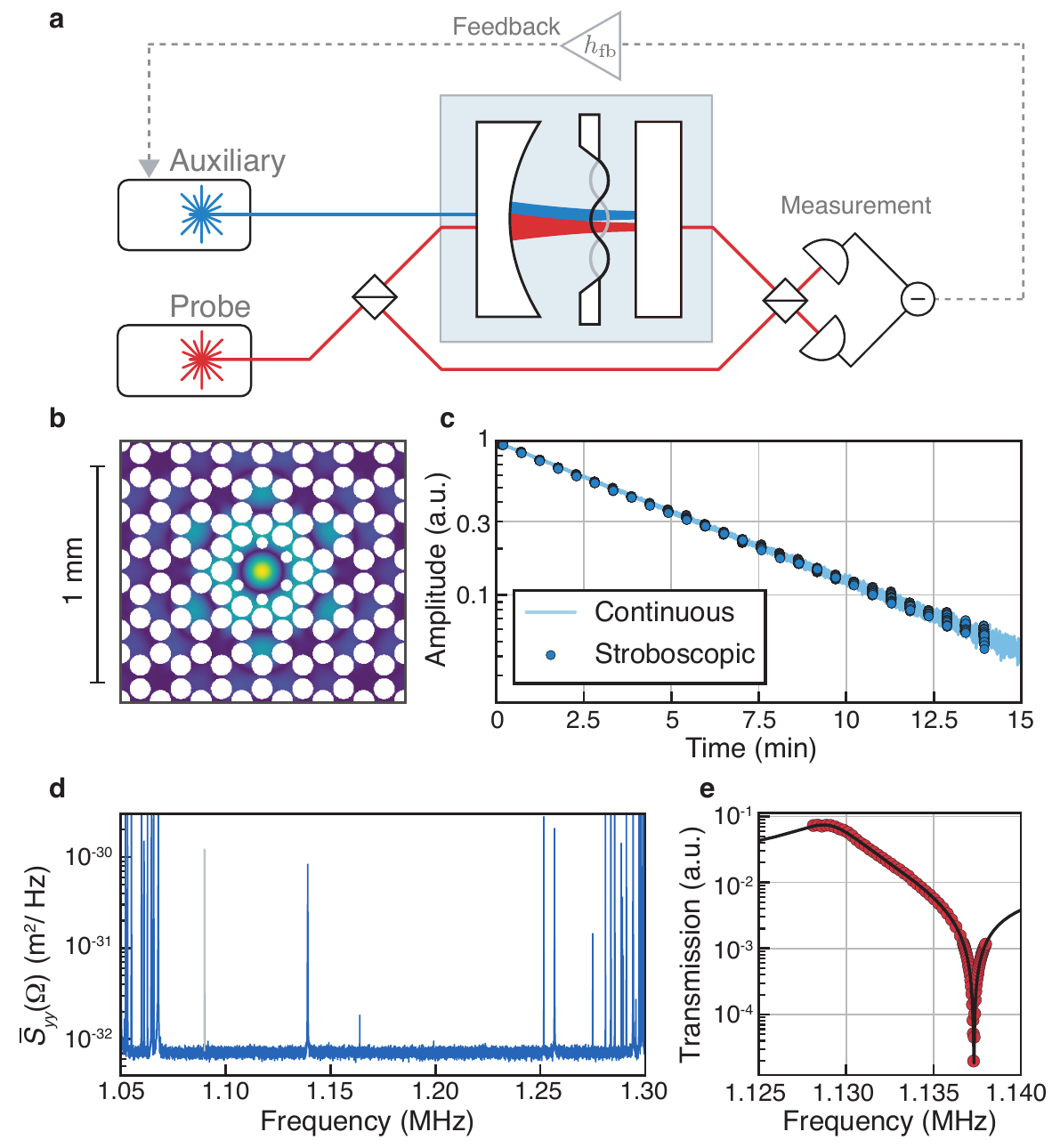}
\caption{{\bf Optomechanical system.} 
{\bf a}, Sketch of experimental setup.
{\bf b}, Simulated displacement pattern (color code) of the mechanical mode of interest.
{\bf c}, Mechanical ringdown measurements.  Light (dark) blue indicates a continuously-monitored (stroboscopic) ringdown. 
{\bf d}, Displacement spectrum around the frequency region of the bandgap.  Out-of-bandgap modes are visible at the edges of the spectrum, and 5 in-gap modes are visible. A phase calibration tone is shown in gray. 
{\bf e}, An optomechanically induced transparency (OMIT) measurement (red symbols), used to characterize the optomechanical coupling strength (from fit, black line).
\label{f:system}}
\end{center}
\end{figure}

\paragraph{Quantum backaction in sideband cooling}
For further characterisation, and direct comparison with a coherent control technique, we proceed with a 
 sideband cooling experiment. 
While monitoring the mechanical motion with a weak ($\Cq\ll1$), resonant probe beam, we lock the auxiliary laser (Fig.~1 with $\hfb=0$) at a finite detuning ($\Daux/2\pi=-$~4.2~MHz).
Increasing the power of this auxiliary beam results in two competing processes: cooling of the motion by optical damping (rate $\Gopt$), and heating by quantum backaction (radiation pressure quantum noise), on top of the constant thermal noise.  
At sufficiently high powers, these processes equilibrate\cite{Aspelmeyer:2014aa, Bowen:2016aa, Peterson:2016aa,Clark:2017aa}, and the mechanical phonon occupancy $\bn=(\Gopt \bnmin+\Gm \bnth)/(\Gopt+\Gm)$ asymptotes to the (sideband cooling) quantum backaction limit $\bnmin=\left((\Om+\Daux)^2+(\kaux/2)^2\right)/(-4 \Daux \Om)$.

Comparison of this model with our data (Fig.~2) allows several interesting conclusions. First, Fig.~2b confirms that the regime of dominating quantum backaction can indeed be deeply accessed, as $\Gopt \bnmin \gg \Gm \bnth$.
Second, the quantum backaction limit ($\bnmin=2.64$) precludes sideband-cooling to the ground state given the ``bad cavity'' ($\kaux\approx \kappa\gg\Om$) employed.
Third, the excellent agreement, even towards the highest $\Gopt$, indicates the absence of significant excess backaction, such as classical radiation pressure noise. 
This is consistent with independent measurements of the lasers' noise (see Supplementary).
Fourth,  equilibration to an optical bath is beneficial for the calibration of the  vacuum optomechanical coupling rate $\vcr$ {\it of the probe} with a standard frequency modulation technique\cite{Gorodetksy:2010aaa} (see Supplementary), which requires reference data with known phonon occupation $\bn$.
Indeed the (usually difficult to ascertain) temperature $T\approx\bnth\,{\hbar}\Om/\kB$ of the phonon thermal bath is insignificant for the largest $\Gopt$; it contributes only 
$(1+\Gopt \bnmin/\Gm \bnth)^{-1}\sim4\%$  to the occupation $\bn$.
Instead, $\bn\approx \bnmin$ is determined by the parameters $\kaux$, $\Daux$ and $\Om$ only, which can be easily and robustly determined spectroscopically.
In physical terms, this means that we use vacuum fluctuations as a temperature reference\cite{Purdy:2017aa} to extract $\vcr$.
Conveniently, a fit (see Supplementary) to the whole dataset with all $\Gopt$, based on standard theory of optomechanical sideband cooling\cite{Aspelmeyer:2014aa}, yields both $\vcr=2\pi\times(127\pm2)\,\mathrm{Hz}$ and $T=(11\pm2) \,\mathrm{K}$.
This compares well to the value of $\vcr=g/\sqrt{\bncav}=2\pi\times129^{+2}_{-3}\,\mathrm{Hz}$ determined from an OMIT fit with a calibrated intracavity photon number $\bncav$.  
Both methods are subject to  different systematic uncertainties (see Supplementary): their excellent agreement underscores a thorough understanding of our system, and lends further support to the crucial calibration---of measured spectra in terms of number of quanta---based on this value of $\vcr$.

\begin{figure}
\renewcommand{\figurename}{{\bf Fig.}}
\begin{center}
\includegraphics[scale=1]{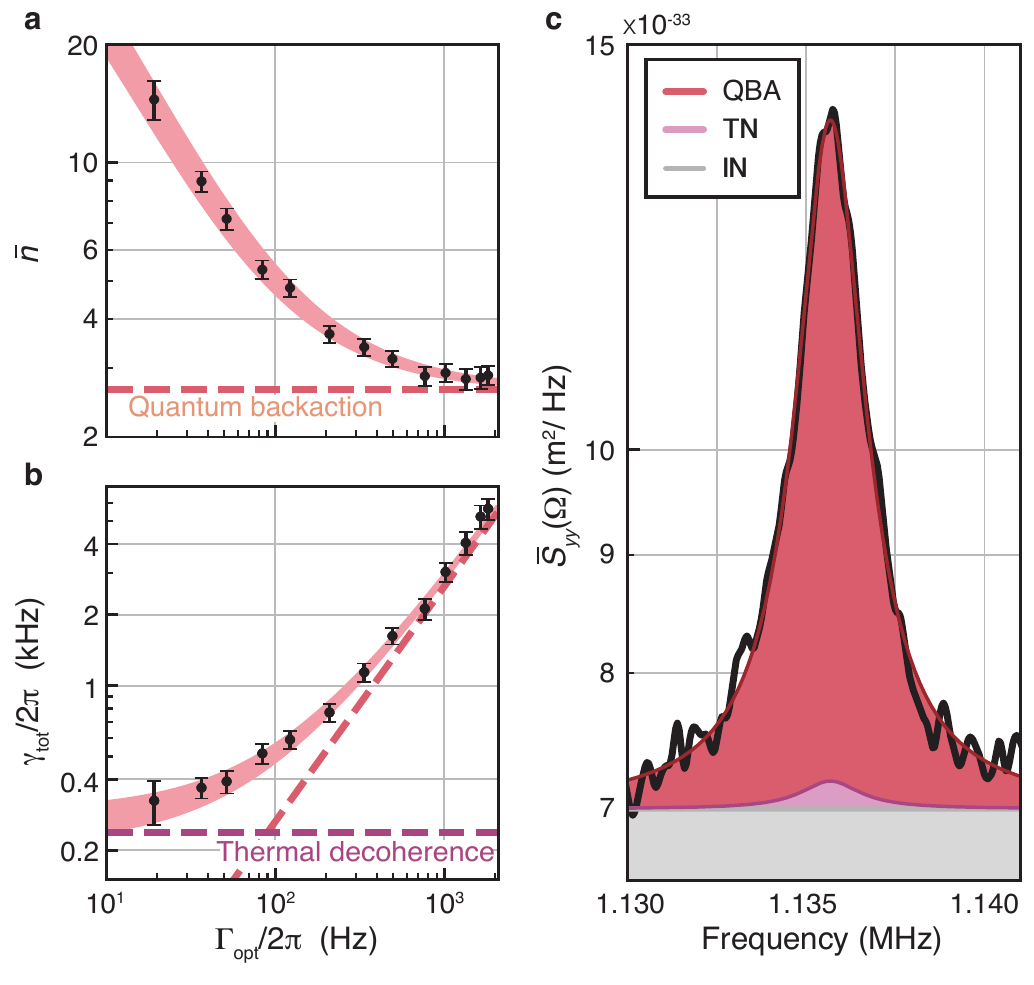}
\caption{{\bf Quantum backaction in sideband cooling.} 
{\bf a, b}, Phonon number $\bn$ and total heating rate $\gtot\equiv \bn(\Gopt+\Gm)=\Gopt \bnmin+\Gm \bnth$ obtained from fitting calibrated displacement spectra ($\Gm$ mechanical energy decay rate, $\Gopt$ optical damping rate, $\bnmin$ sideband cooling limit occupancy, $\bnth$ thermal bath occupancy).
Solid lines are fits, and the thickness reflects the uncertainty in the fit parameters.
Dashed lines indicate contributions from thermal decoherence (purple) and quantum backaction (red). Error bars indicate standard deviation.
{\bf c}, Calibrated displacement spectrum corresponding to the highest cooling power with Lorentzian fit.
The gray line represents the imprecision noise (IN).
Thermal noise (TN) contributes $\sim 4\%$ of the total force noise, the remainder being due to quantum backaction (QBA).
\label{f:quantumBackaction}}
\end{center}
\end{figure}
\paragraph{Quantum measurement}
In the next experiment we characterize the quality of the measurement, to gauge the possibility of overcoming the sideband cooling limit via measurement-based quantum feedback with $\eta~\approx~1$.
To this end, we reduce the auxiliary laser power and arrange it to provide only mild precooling on the mechanical mode of interest ($\Gopt/2\pi\sim \mathcal{O}(10)\,\mathrm{Hz}$), as well as all other modes of the membrane (see Supplementary), to stabilize the system.
At the same time, we increase the probe power in several steps into the regime $\Cq>1$.

Figure~3 shows the corresponding measured mechanical displacement spectra $\bar{S}_{yy}(\Og)$, obtained from the calibrated homodyne photocurrent $y(t)=x(t)+x_\mathrm{imp}(t)$, which contains both the actual mechanical position ($x$) and the measurement imprecision noise ($x_\mathrm{imp}$) (see Supplementary). 
When increasing the probe strength, the imprecision noise floor decreases, and the total force noise on the resonator increases due to quantum backaction. 
We fit the spectra to a Lorentzian peak, driven by a total force noise $\SFFtot$, and with an imprecision noise floor $\Sxximp$, 
\begin{equation}
  \bar S_{yy}(\Og)=|\chieff (\Og)|^2 \SFFtot+\Sxximp.
  \label{e:Sxx}
\end{equation}
Here $\chieff (\Og)$ is the effective mechanical susceptibility, with resonance frequency $\Oeff$ and damping $\Geff$, affected by the lasers' (largely irrelevant) dynamical backaction\cite{Aspelmeyer:2014aa,Bowen:2016aa}.

For the sake of comparability, we can reference these measurement noises to the resonant spectral density associated with mechanical zero-point fluctuations.
This yields the number of imprecision noise quanta  $\nimp=\Sxximp/(8 \xzpf^2/\Gm)$ and force noise quanta $n_\mathrm{tot}=\SFFtot/(8 \pzpf^2 \cdot \Gm)$, where $\pzpf$ is the momentum zero-point-amplitude such that $\xzpf \pzpf=\hbar /2$.
For the strongest measurement we find $\nimp=5.8\times 10^{-8}$.
This constitutes a three-order of magnitude improvement over the best measurements to-date\cite{Wilson:2015aa}.
All measured values agree with the expectation $\nimp= \Gm/16 \Gmeas$ to within a factor of $1{.}03\pm 0{.}06$.
The force acting on the system can be broken down into three contributions, $ \SFFtot=\SFFth+\SFFaux+\SFFqba $, due to thermal noise and quantum backaction of the auxiliary and probe beam, respectively.
For each data set, the obtained fit results agree with the $\SFFtot$ predicted  from the system's parameters (see Supplementary) to within a factor of $1{.}08\pm0{.}02$, whereby $\SFFaux/\SFFth\approx 0{.}18$  and $\SFFqba/\SFFth=\Cq$.

We can use these findings to evaluate the probe's measurement efficiency, $\eta=(16\nimp n_\mathrm{tot})^{-1}=56\%$, on par with circuit QED systems\cite{Vijay:2012aa} and sufficient to exert high-fidelity quantum control. 
Using $\eta=\hbar^2/(\Sxximp \SFFtot)$ further allows comparison to the Heisenberg measurement-disturbance uncertainty relation\cite{Braginsky:1992aa,Aspelmeyer:2014aa,Bowen:2016aa}
\begin{equation}
  \sqrt{\Sxximp \SFFqba}\geq\hbar.
  \label{e:Heisenberg}
\end{equation}
The measured {\it total} noises $1{.}33 \, \hbar=\sqrt{\Sxximp \SFFtot}\geq\sqrt{\Sxximp \SFFqba}$, then constrain the deviation from an ideal measurement to at most $33 \%$.
To our knowledge, this is the closest mechanical realization of the Heisenberg microscope Gedankenexperiment to date.
Consequently, the experimental displacement sensitivity of equation~(\ref{e:Sxx}) also approaches the SQL for such measurements more closely than ever before.
Indeed, we find that off the mechanical resonance ($\delta \Omega=\Og-\Oeff \approx 2\pi\times3.3\,\mathrm{kHz}$), where the uncorrelated imprecision and backaction noises are optimally balanced, our mechanical sensor reaches $\bar S_{yy}(\Oeff+\delta \Omega)=1{.}35\,\bar S_{yy}^\mathrm{SQL}(\Om+\delta \Omega) $, where $\bar S_{yy}^\mathrm{SQL}(\Og)=2\hbar |\chim(\Og)|$.
This is better than what is currently achievable in advanced LIGO\cite{Martynov:2016aa}, with ultracold atoms\cite{Schreppler:2014aa}, or with ultracold mechanical resonators\cite{Teufel:2011aa}, even when probed with squeezed light\cite{Clark:2017aa} or with nominally sub-SQL variational techniques\cite{Kampel:2017aaa}.

\begin{figure}
\renewcommand{\figurename}{{\bf Fig.}}
\begin{center}
\includegraphics[width= 0.99\linewidth]{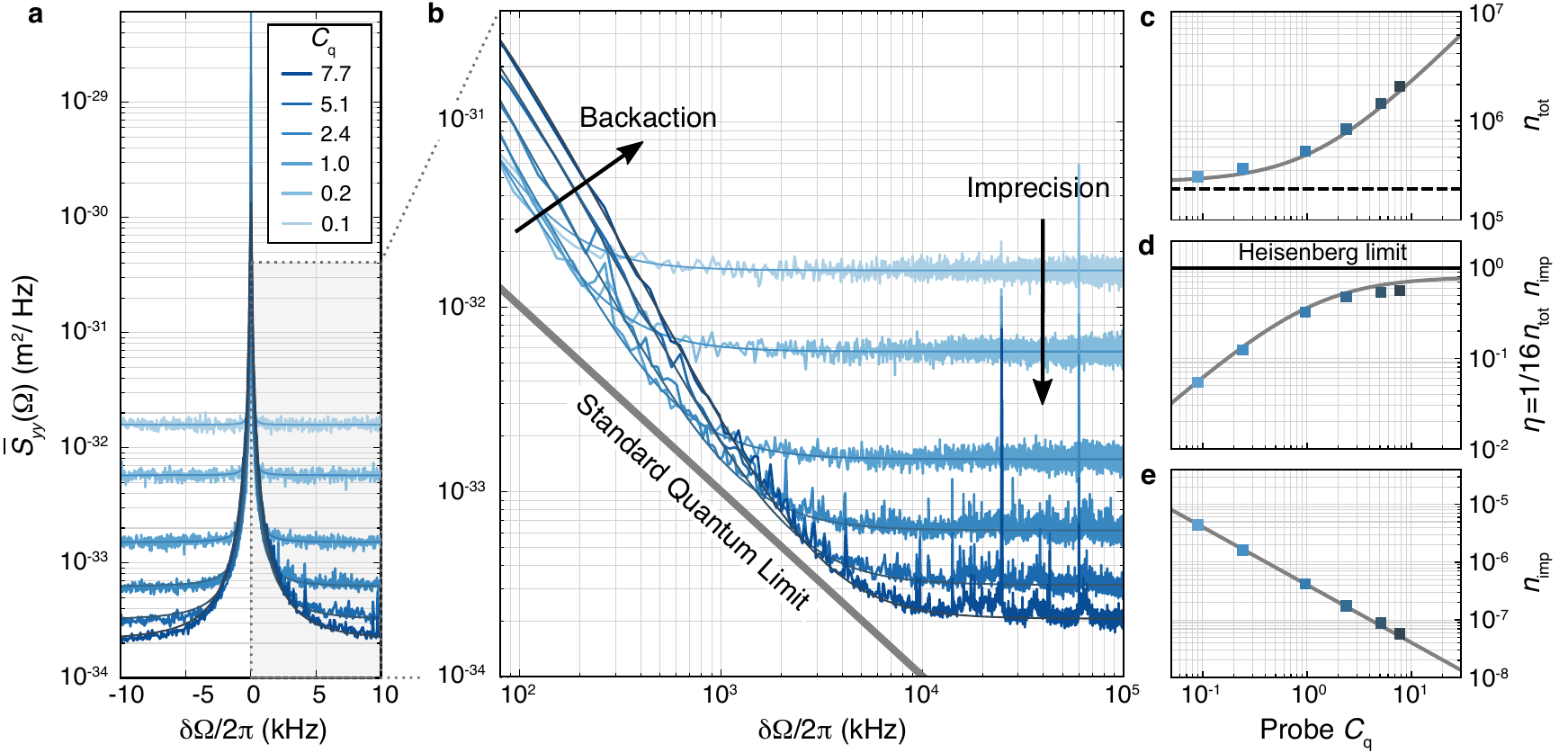}
\caption{{\bf Quantum measurement.} 
{\bf a}, Displacement spectra around the effective mechanical frequency $\Oeff$, for different probe quantum cooperativity $\Cq$.
{\bf b}, Off-resonant tails of the spectra from {\bf a}.
Lorentzian fits to the spectra yield imprecision $n_{\mathrm{imp}}$ and total force noise $n_{\mathrm{tot}}$. They are shown, as a function of quantum cooperativity $\Cq$, in {\bf c, e}, respectively. The former is visible as a decrease of the noise floor, while the latter appears as a rise in the wings of the Lorentzian.
From these values, one can calculate the measurement efficiency, $\eta=1/16\nimp n_\mathrm{tot}$, which reaches 0.56, shown in {\bf d}.  Note that the Heisenberg limit corresponds to ideal efficiency, $\eta=1$. 
\label{f:quantumMeasurement}}
\end{center}
\end{figure} 
\paragraph{Ground state cooling by feedback}
We now use the signal $y(t)$ obtained from this near-ideal quantum measurement to control and stabilize the quantum state of the mechanical system.
To this end, we electronically convolve it with a filter kernel $\hfb(t)$ and apply the output $\Ffb(t)=\hfb(t)*y(t)$ as a force to the mechanical resonator (feedback interaction $  H_\mathrm{fb}=\Ffb(t) \hx(t)$). To exert this force we modulate the amplitude of the auxiliary laser beam, whose power is kept small as in the previous  experiment.
In the domain of linearized Quantum Optomechanics, assuming Gaussian noises only, the system's quantum dynamics can be conveniently mapped to a classical control problem, with the important caveat that process and measurement noises must be included that mimic the quantum-mechanically required backaction and imprecision, respectively.
Linear-Quadratic-Gaussian (LQG) control theory then provides a straightforward path to obtain the optimum controller for cooling, whose objective is to reduce a quadratic cost function, namely the mechanical position and momentum variance of a single mechanical mode\cite{Wiseman:2009aa,Jacobs:2014aa,Bowen:2016aa,Doherty:1999aa,Doherty:2000aa,Garbini:1996aa} (see Supplementary).

While inspired by these results, our feedback filter $\hfb(\Og) = \hfbmain(\Og)+\hfbaux(\Og)$, with
\begin{equation}
  \hfbmain(\Og) = \gfb e^{\imath\Og\tau - \imath\phi}\left[\frac{\Gammafb\Og}{\Ogfb^2-\Og^2-\imath\Gammafb\Og}\right]^2,
  \label{e:filterMain}
\end{equation}
accommodates a more complex experimental reality (see Supplementary).
In particular, it contains a predominantly electronic loop delay $\tau\approx 300\,\mathrm{ns}$, a high order bandpass filter (bandwidth $\Gammafb$, gain $\gfb$, global phase $\phi$) peaked at the center frequency $\Ogfb$, close to the phononic bandgap's center, to suppress gain for out-of-gap modes, as well as an auxiliary filter $\hfbaux$ that suppresses instabilities of other mechanical modes far off $\Om$.
The phase $\phi$ is electronically adjusted such that $\arg(\hfb(\Om))\approx -\pi/2$.
The feedback force is then approximately proportional to the resonator's velocity, providing a quantum-noise limited friction force, sometimes referred to as ``cold damping''\cite{Mancini:1998aa,Cohadon:1999aa,Genes:2008ab}. 
Together with standard optomechanical theory\cite{Aspelmeyer:2014aa,Bowen:2016aa}, equation~\eref{e:filterMain} can be incorporated into a simple control-theoretical model which predicts the  spectra $\bar{S}_{yy}(\Og)$ of the measured displacement, and the underlying resonator position and momentum fluctuations $\bar{S}_{xx}(\Og)$, $\bar{S}_{pp}(\Og)$, respectively (see Supplementary).

To  assess the cooling performance, we fit the predicted $\bar{S}_{yy}(\Og)$ to measured spectra, adjusting $n_{\mathrm{tot}}$, $n_{\mathrm{imp}}$, feedback gain ($ \gfb$) and phase ($\phi$).
The fit values for $n_{\mathrm{tot}}$ and $n_{\mathrm{imp}}$ agree with independent, first-principles calculations to within factors of $1{.06}\pm0{.}07$ and $1{.01}\pm0{.}05$, respectively.
We then proceed to calculate the occupation of the mechanical resonator from its position and momentum variance, $\bn=\langle \hbd \hb \rangle\approx\left( \int_0^\infty \bar{S}_{xx}(\Og)\xzpf^{-2} d\Og/2\pi-1\right)/2$.
Figure~4 shows the results as a function of controller gain (expressed as effective resonator damping $\Geff=\Gm+\mathrm{Im} \left[h_{\mathrm{fb}}(\Om)\right]/m\Om+\Gopt$), for five different probing strengths up to $ \Cq =$~7.8.
For each $\Cq$, a minimum occupancy is reached for a certain gain, beyond which the resonator is heated again, a mechanism known as noise ``squashing''\cite{Cohadon:1999aa,Poggio:2007aa,Wilson:2015aa} (Fig.~4b), in which significant imprecision noise is fed back to the mechanics.
The lowest residual occupation observed is $\bar{n} = 0.29\pm0.03$ (Fig.~4a).

\begin{figure}[t]
\renewcommand{\figurename}{{\bf Fig.}}
\begin{center}
\includegraphics[scale=1]{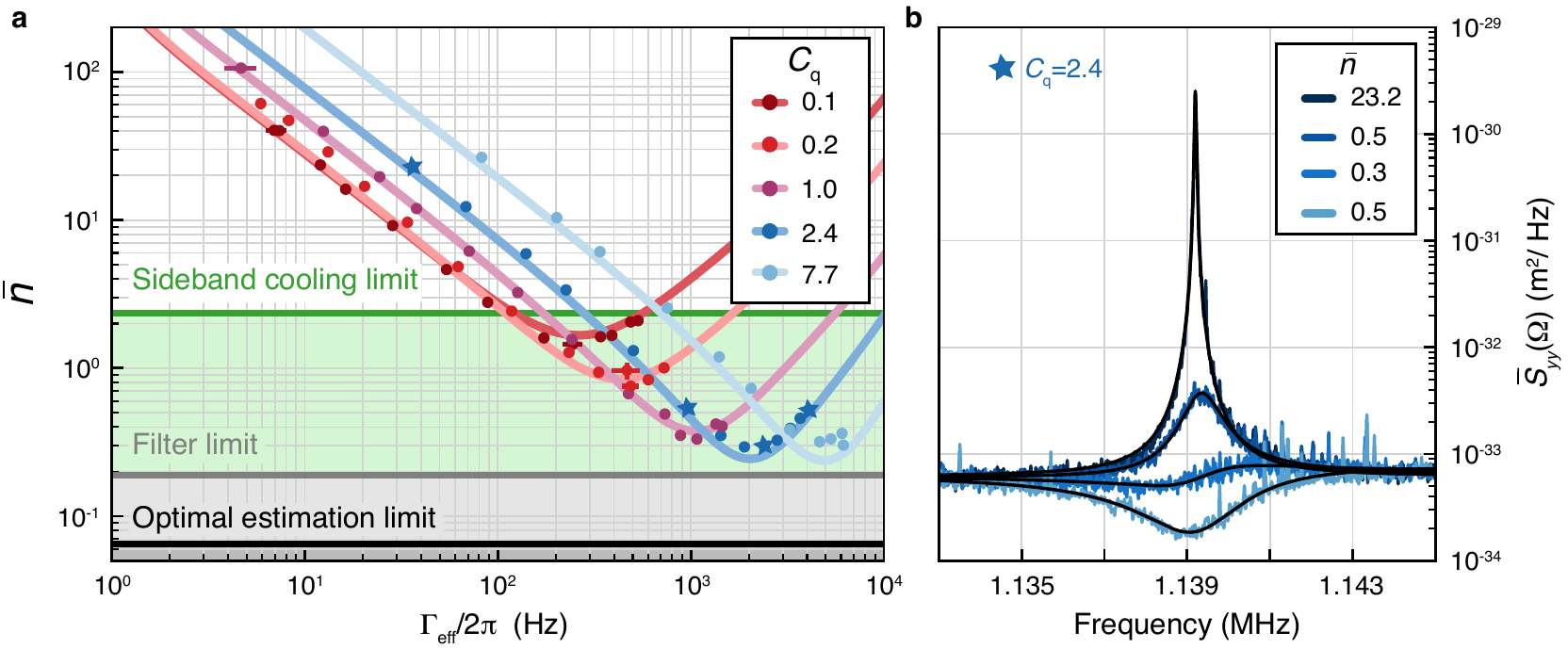}
\caption{{\bf Feedback cooling to the quantum ground state.}
{\bf a}, Mechanical occupancy for different values of quantum cooperativity $\Cq$, as a function of loop gain, expressed here as an effective damping rate $\Geff$.
Points are data, where error bars indicate fit uncertainty; 
solid lines are theoretical calculations using independently-estimated system parameters.
The sideband cooling limit for this system is shown as a green line.  Gray and black lines indicate the limit of our implemented filter, and of optimal state estimation, respectively (see Supplementary).
{\bf b}, Exemplary spectra for $\Cq=$~2.4, at the gain values marked with a star in {\bf a}.
\label{f:feedbackCooling}}
\end{center}
\end{figure}

We can benchmark this cooling performance against the occupation $\bnest$ of the conditional state (i.e., conditioned on the measurement result)\cite{Doherty:2012,Bowen:2016aa,Wieczorek:2015aa}.
The lowest conditional occupation is reached by optimal state estimation from the available measurement record, and coincides with the lowest occupation to which ideal feedback can bring the resonator.
Under the idealized assumption of a single-mode, high-Q resonator coupled to a hot thermal bath\cite{Doherty:2012,Bowen:2016aa}
\begin{equation}
	\bnest\approx \frac{1}{2} \left(\sqrt{\frac{1}{\eta}}-1 \right)
\end{equation}
yielding $\bnest\approx$~0.07 for our $\etadet=$~0.77 and $\Cq\rightarrow \infty$, corresponding to a state of high purity  ($(1+ \bnest)^{-1}\approx 93\,\%$).
The discrepancy with the achieved occupation indicates further room for improvement in engineering our feedback filter, whose sub-optimal nature becomes apparent for high quantum cooperativity and gain (see Supplementary).
The performance compromise is a consequence of the need to avoid instabilities anywhere outside the bandgap---spectral regions crowded with other high-quality mechanical modes.
Wider bandgaps, reduced loop delay, and a filter accounting for individual out-of-band modes, are thus obvious routes to improved feedback.

By turning off the electronic feedback abruptly after cooling close to the ground state, we can directly measure the resonator's heating rate.
Figure~5 shows the result of such a measurement, averaged over $\sim400$ experimental iterations.  
The occupancy $\bn(t)$  equilibrates exponentially to the level given by residual sideband cooling.  
At low probe power, we infer (see Supplementary) a heating rate of $1{.}4$~phonons/ms {out of the ground state} from the slope of this curve at $t=0$~ms.
This is consistent with the expected thermal decoherence rate $\gamma\approx\bnth \Gm $ at this experiment's $T\approx 9\,\mathrm{K}$  plus a $0{.}2$~phonons/ms contribution due to quantum backaction.

\begin{figure}
\renewcommand{\figurename}{{\bf Fig.}}
\begin{center}
\includegraphics[scale=1]{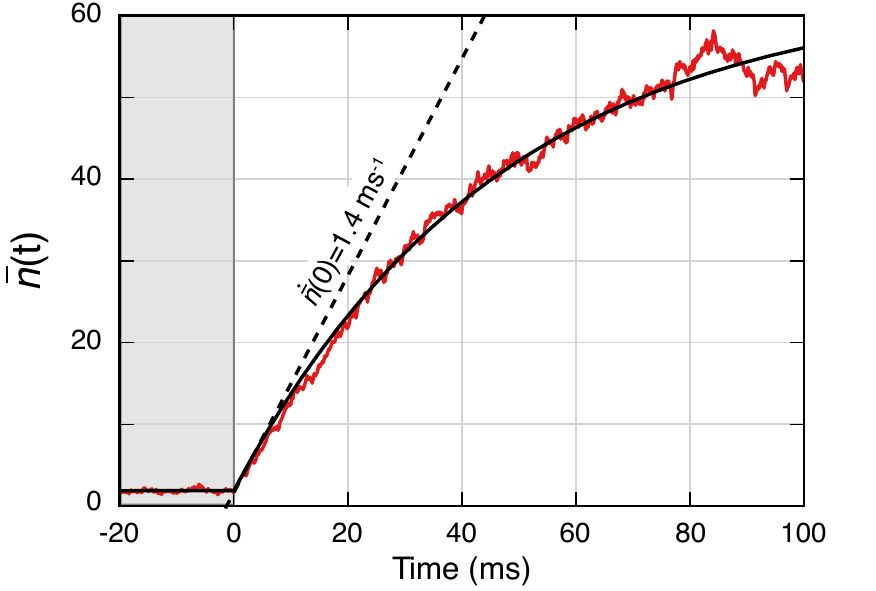}
\caption{{\bf Heating from low phonon occupancy.}
Mechanical heating $\bn(t)$ (red trace) measured by switching off the feedback at $t=0$~ms. Black line is an exponential fit, its dashed tangent indicates $\dot \bn(t=0)\approx\gamma_\mathrm{tot}$.
\label{f:heatingRate}}
\end{center}
\end{figure}
\paragraph{Discussion and outlook}
Such ms-coherence times compare favorably with other mechanical systems held at dilution refrigerator temperatures\cite{Peterson:2016aa,Clark:2017aa,Chu:2017aa}, and might be boosted to seconds if made similarly cold.
The very narrow linewidth ($\sim1\,\mathrm{mHz}$) of the mechanical mode furthermore suggests the absence of significant dephasing (see Supplementary).
Quantum memory applications are thus conceivable.

With an overall measurement efficiency close to unity, mechanical systems such as the one presented here will allow test and application of a wide range of  quantum measurement and control techniques\cite{Wiseman:2009aa,Jacobs:2014aa,Bowen:2016aa}.
This includes time-continuous Bell measurements for state teleportation and entanglement swapping\cite{Hofer:2013aa,Hofer:2015aa} and  a combination with mechanical parametric amplification\cite{Szorkovszky:2011aa} to create strongly squeezed mechanical states.
The achieved occupancy 9~dB below the sideband cooling limit is promising for quantum control, in particular, of low-frequency mechanical systems, such as those employed for gravitational wave detection.
Indeed, feedback based protocols to enhance interferometric detectors have already been proposed\cite{Courty:2003aa}.
Combination with nonlinear measurement schemes, such as photon counting\cite{Ringbauer:2018aa}, could allow non-Gaussian state preparation.
Moreover, the fascinating prospect of quantum control of motion at room temperature also appears more realistic than ever, given that $\Cq>100$ is achieved at $10\,\mathrm{K}$.
 
\clearpage
\newpage

 {\bf Acknowledgements} The authors acknowledge discussions with K.~Hammerer, E.~Zeuthen, D.~Vitali and sample fabrication at early stage from Y.~Seis.
This work has received funding from the European Union's Horizon 2020 research and innovation programme (European Research Council (ERC) project Q-CEOM, grant agreement no.~638765 and FET proactive project HOT, grant agreement no.~732894), a starting grant from the Danish Council for Independent Research, and the Carlsberg Foundation.
\clearpage
\newpage

\renewcommand{\figurename}{{\bf Supplementary Fig.}}
\renewcommand{\tablename}{{\bf Supplementary Table}}
\setcounter{figure}{0}\renewcommand{\thefigure}{{\bf S\arabic{figure}}}
\setcounter{table}{0}\renewcommand{\thetable}{{\bf S\arabic{table}}}
\setcounter{equation}{0}\renewcommand{\theequation}{S\arabic{equation}}
\newgeometry{top=40mm, bottom=20mm, right=20mm, left=20mm}
\baselineskip18pt

\vspace{3mm}

\hspace{-6mm}{\LARGE \bf Supplementary Information}

\vspace{5mm}

\tableofcontents

\newgeometry{top=24.5mm, bottom=26.8mm, right=18mm, left=17mm}
\baselineskip18pt
\newpage
\subsection*{System parameters}\label{sec:sys_parms}
\addcontentsline{toc}{section}{System parameters}

\begin{table}[h]
\centering
\resizebox{0.72\textheight}{!}{\begin{tabular}{llll}
    \hline
    Symbol & Definition & Name & Value  \\ \hline
	$\Og$ & & Fourier frequency \\
    $\Om$ & & Mechanical resonance frequency & $2\pi\times$~1.139~MHz \\
    $\Qm$ & $\Om/\Gm$ & Mechanical quality factor & $1.03^{+0.03}_{-0.02}\times10^9$ \\
    $\Gm$ &  & Mechanical damping rate & $2\pi\times$~1.09~mHz \\
    $\meff$ & & Effective mass & 2.3~ng \\
    $\xzpf$ & $\sqrt{\frac{\hbar}{2\meff\Om}}$ & Displacement zero point fluctuations & 1.8~fm \\
    $\Tm$ & & Mechanical bath temperature & 11~K \\
    $\vcr$ & & Vacuum optomechanical coupling, probe & $2\pi\times$~127~Hz\\
    $\kpr$ & & Linewidth of probe cavity mode & $2\pi\times$~15.9~MHz\\
    $\Dpr$ & & Detuning of probe laser & $ -0.05\,\kpr<\Delta<0$\\
    $\bncavpr$ & & Intracavity photons, probe laser & \\
    $\etac$ & & Cavity outcoupling, probe laser & 0.95 \\
    $\vcrAux$ & & Vacuum optomechanical coupling, auxiliary  & \\
    $\kaux$ & & Linewidth of auxiliary cavity mode & $2\pi\times$~12.9~MHz\\
    $\Daux$ & & Detuning of auxiliary laser & \\
    $\bncavaux$ & & Intracavity photons, auxiliary laser  & \\
    $\eta_\mathrm{c,aux}$ & & Cavity outcoupling, auxiliary laser & 0.88 \\
    $\eta_\mathrm{det}$ & & All-in homodyne detection efficiency & 0.77 \\    
    $\tau$ & & Feedback loop delay & 300~ns  \\
    $\Ogfb$ & & Center of electronic bandpass filter & $2\pi\times$~1.1925~MHz \\
    $\Gammafb$ & & Bandwidth of electronic bandpass filter & $2\pi\times$~77.79~kHz \\
     & & & \\
    $\chim(\Og)$ & $\meff^{-1}\left(\Om^2 - \Og^2 - \imath\Og\Gm\right)^{-1}$ & Mechanical susceptibility & \\ 
    $\hfb(\Og)$ &  & Feedback filter transfer function & \\
    $\bar S_{vv}(\Og)$ & & Symmetrized, single-sided & \\
     & & power spectral density of variable $v(t)$ & \\
     $x(t)$ & & Mechanical displacement & \\
     $\ximp(t)$ & & Measurement imprecision noise & \\
     $y(t)$ & $x(t)+\ximp(t)$ & Measured mechanical displacement & \\ 
     $\Gamma_\mathrm{meas}$ & $4\eta_\mathrm{det}g^2/\kappa$ & Measurement rate & \\
     $\Gamma_\mathrm{qba}$ & $4g^2/\kappa$ & Measurement-induced quantum backaction& \\
     $\gamma$ & $\Gm \bnth $ & Thermal decoherence & \\
     $\Cq$ & $\Gamma_\mathrm{qba}/\gamma$ & Quantum cooperativity & \\
     $\eta$ & $\Gmeas/(\Gamma_\mathrm{qba}+\gamma)$ & Measurement efficiency &  \\\hline
\end{tabular}}
\caption{Parameters and definitions}
\label{tab:parameters}
\end{table}

\clearpage
\newpage
\section{Experimental details}
\subsection{Soft-clamped mechanical resonator}\label{sec:ultra_membrane}
The mechanical device used in the experiment is based on a (20nm$\times$3.6mm$\times$3.6mm) soft-clamped $\mathrm{Si_3N_4}$ membrane\cite{Tsaturyan:2017aa}. As shown in Supplementary~Fig.~S1a, a honeycomb hole pattern is fabricated into the membrane, producing phononic bandgaps for out-of-plane modes. In the center of the membrane, a defect is created, supporting localized vibrational modes whose frequencies lie in one of these bandgaps. These mechanical modes are ``soft-clamped", in the sense that their mode shapes decay into the phononic crystal structure gradually, as opposed to being clamped by a rigid frame.  This reduced curvature, combined with stress redistribution due to the phononic pattern, results in ultra-high mechanical quality factors. In this work we focus on mode A, at $\Om/2\pi=$~1.139~MHz. Compared to a previous work in ref.~\cite{Tsaturyan:2017aa}, a modified defect design is used, in order to shift mode A away from the left bandgap edge. 

\begin{figure}[h]
\begin{center}
\includegraphics[scale=1]{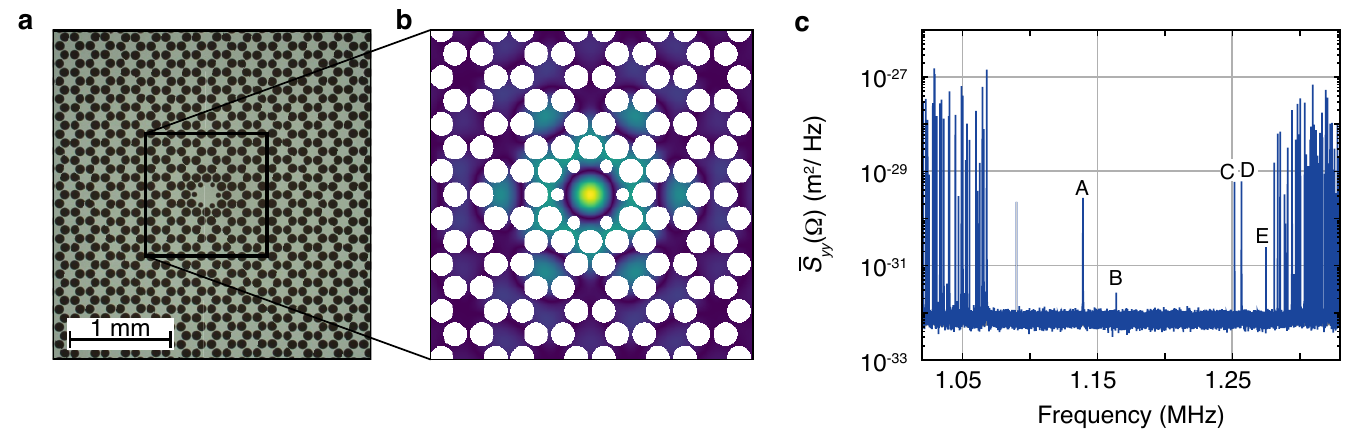}
\caption{{\bf Soft-clamped membrane.}
{\bf a}, Photograph of the soft-clamped membrane.
{\bf b}, Simulated displacement pattern of defect-localized mode A.
{\bf c}, Mechanical spectrum of the lowest-frequency bandgap, with defect-localized modes labeled from A to E. The gray peak at 1.09 MHz is a phase calibration tone.
}
\end{center}
\end{figure}

To measure the quality factor of this soft-clamped mode, ringdown experiments are performed. The laser is tuned to a wavelength where the finesse of the optical cavity is low ($F\sim\mathcal{O}(10)$), allowing interferometric displacement measurements without dynamical optomechanical effects. The transmitted light {intensity} is {directly} measured with a photodiode, and the photocurrent is demodulated at the mechanical frequency $\Om$, to obtain a record of the motion.  To excite a desired mechanical mode, the amplitude of {an auxiliary} laser is modulated at $\Om$. When the modulation is turned off, the oscillation amplitude decays according to $x(t)=x(0)e^{-\Om t/(2 \Qm)}$, as shown in Supplementary~Fig.~S2a. {From a fit we extract a quality factor $\Qm=1.03\times10^9$.}

When measuring such extreme quality factors, it is important to ensure that the decay is not modified by any residual dynamical effects, due to photothermal or radiation pressure backaction.
\begin{figure}[t]
\begin{center}
\includegraphics[scale=1]{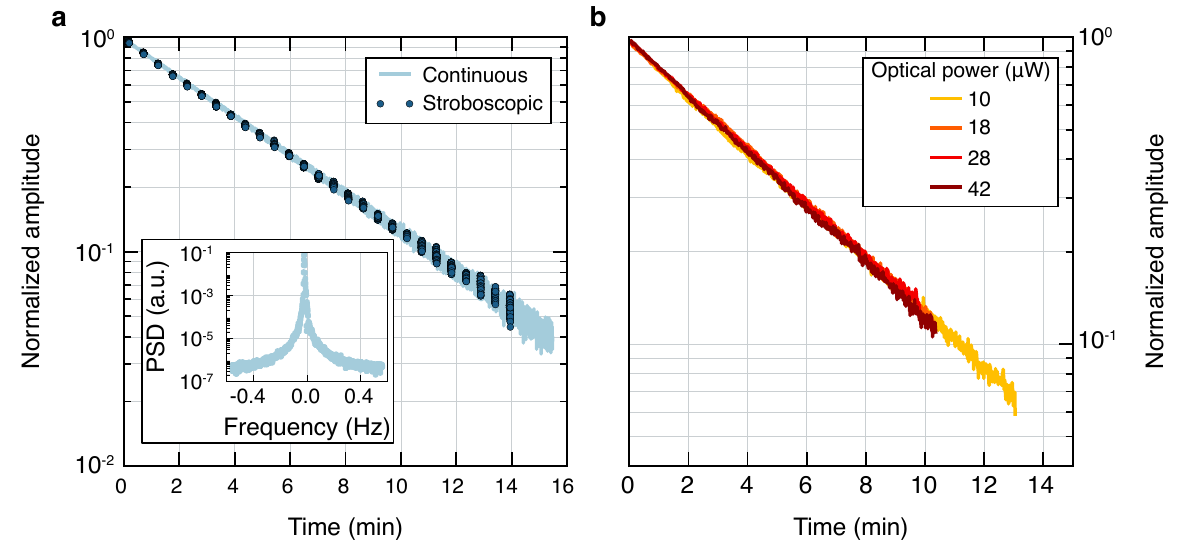}
\caption{{\bf Mode A ringdowns.}
{\bf a}, Ringdowns with continuous and stroboscopic optical monitoring. The inset shows the power spectral density of the continuous ringdown data.
{\bf b}, Ringdowns at different continuous optical powers. The extracted Q values are 1.02, 1.06, 1.07, 1.04 billion from high to low optical power.
}
\end{center}
\end{figure}
In Supplementary~Fig.~S2b, we confirm that the measured Q does not depend on laser power, as would be expected for these effects.
We also conduct a ``stroboscopic'' ringdown measurement, in which the motion was only probed for brief moments (duty cycle $\sim$~4\%, period $\sim$~0.5~min). The continuous and stroboscopic ringdowns overlap well, yielding $Q$-factors of 1.03$\times10^9$ and 1.02$\times10^9$, respectively. The inset of Supplementary~Fig.~S2a shows the power spectral density (PSD) of the continuous ringdown data. The width of this peak is Fourier-limited to 1.1~mHz for these data, and thereby confirms the absence of significant dephasing, since the energy decay rate was found (by ringdown) to be 1.1 mHz.

\subsection{Experimental setup}\label{sec:exp_setup_details}
\begin{figure}[t]
\begin{center}
\includegraphics[scale=1]{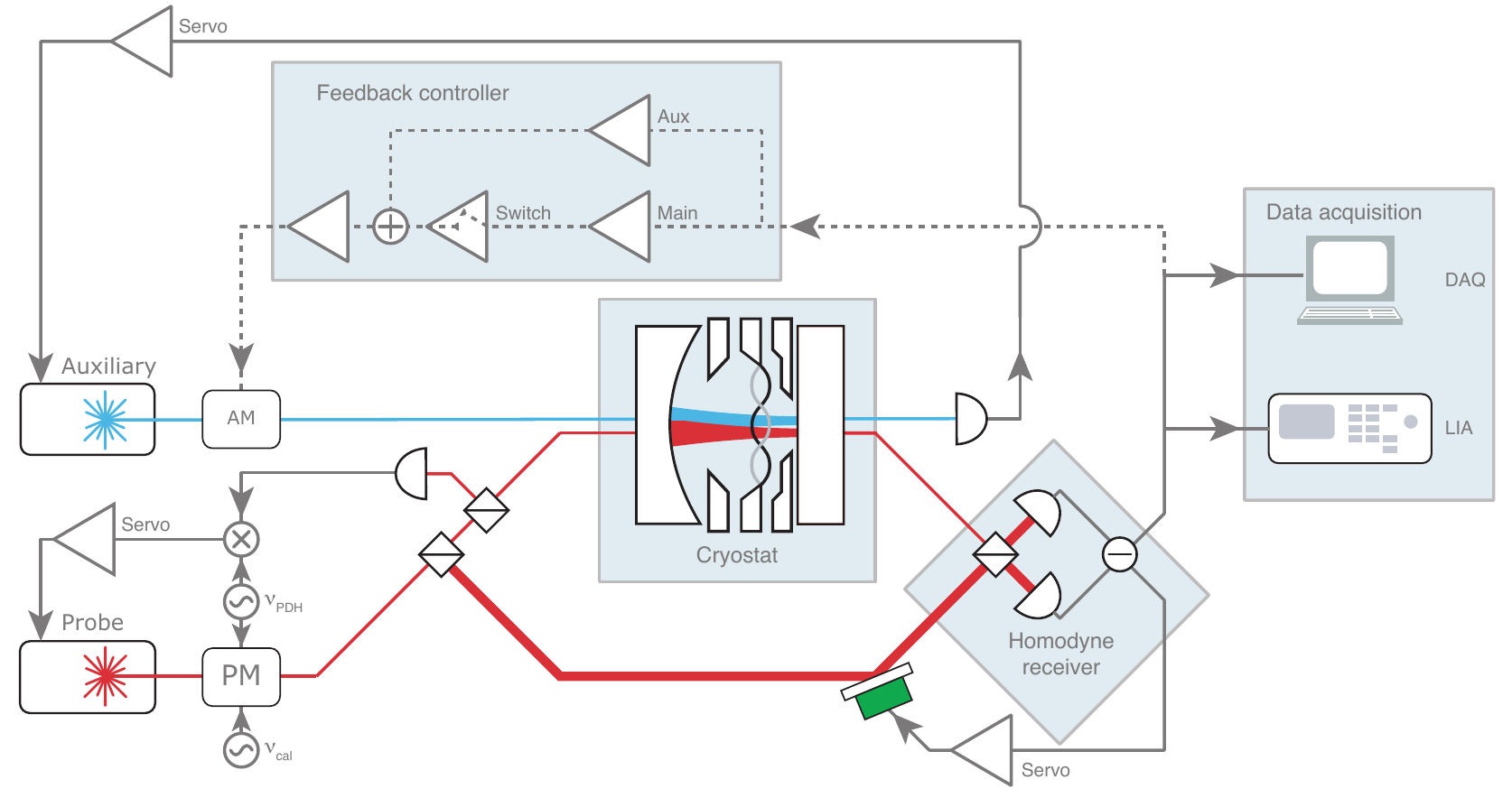}
\caption{{\bf Experimental setup.}
Overview of the optical and electronic scheme used in the experiments.
}
\end{center}
\end{figure}
Supplementary~Fig.~S3 shows more details about the setup for the experiments described in the main text.
A Ti:S laser (red) at $\lambda\approx$~795~nm is used to probe the frequency fluctuations of an optical cavity mode, whose linewidth is $\kpr/2\pi=$~15.9~MHz.
To stabilize the laser frequency relative to the optomechanical cavity, we implement a Pound-Drever-Hall (PDH) scheme\cite{Black:2001aa}, using a phase modulator (PM) on the probe beam. This same PM is also driven with a coherent tone at $\nu_{\mathrm{cal}}$, for calibrating the transduction of optical frequency fluctuations into detected voltage fluctuations\cite{Gorodetksy:2010aa}.
At the wavelength of this probe beam, the reflectivities of the two cavity mirrors differ significantly, forming a strongly asymmetric optical resonator. To detect as much of the cavity light as possible, we drive the cavity through the higher-reflectivity port while detecting the light leaving the more transmissive port. We perform a phase-sensitive measurement on the transmitted light by means of a balanced homodyne receiver. 

To stably measure the signal beam's optical phase, on which the mechanical displacement information is imprinted, we stabilize the path difference of the signal and local oscillator arms.  A feedback loop actuates a piezo-controlled mirror in the local oscillator path, minimizing the DC component of the photocurrent \cite{Leonhardt:1997aa}.
The information about the mechanics is contained in the AC part of the photocurrent, which is digitally acquired both from a data acquisition card (DAQ) to perform a Fourier analysis and from a digital lock-in amplifier (LIA) to analyze the time evolution (see Sec.~\ref{sec:si_heating_rate}).

An auxiliary Ti:S laser (blue) at $\lambda_\mathrm{aux}\approx$~796~nm is frequently used in the experiment. 
To avoid unwanted interference, its polarization is orthogonal to the probe laser, and it is locked to a different longitudinal cavity mode, whose linewidth is $\kaux/2\pi =$~12.9~MHz.
In the experiment described in Fig.~2, this laser provides the sideband cooling and acts as a source of strong quantum backaction.
In the feedback cooling experiment (Fig.~4), the auxiliary laser is used (in combination with an amplitude modulator) to exert a force on the mechanical resonator via radiation pressure, i.e. to actuate the feedback force.
 For this feedback, an FPGA-based digital controller (RedPitaya 125-14) \cite{RedPitaya:2016aa} is used to band-pass filter the AC homodyne photocururent close to the mechanical mode to be cooled. The filter implementation is implemented on an open-source, Python-based software module\cite{Neuhaus:2017aa}, whereby the built-in I/Q modulation capability enables filters with continuously tunable phase\cite{Neuhaus:2016aa}. The processed signal is amplified and sent to a fiber-integrated optical amplitude modulator on the auxiliary beam.
In the actual experiment we employ several FPGA controllers. One of them is devoted to cool the defect mode of interest, employing a transfer function $\hfbmain(\Og)$ given by equation~(4). 
An electronic switch is inserted just after this controller to toggle on and off the feedback force to measure transient dynamics. 
All  other controllers can be grouped in a single transfer function $h_\mathrm{aux}(\Og)$, used for cooling some of the low frequencies modes corresponding to motion of the entire $\mathrm{Si_3N_4}$ membrane structure, as well as defect modes C and D. This is needed to avoid large frequency fluctuations and keep the whole system stable.
This auxiliary controller is always on in the experiments reported in Fig.~3 and Fig.~4.

\subsection{Optomechanical assembly}\label{sec:om_assembly}

We largely use the same optomechanical assembly and optical characterization techniques as described in ref.~\cite{Nielsen:2017aa}, combining it here with a soft-clamped membrane\cite{Tsaturyan:2017aa}.
In this membrane-in-the-middle geometry\cite{Thompson:2008aa}, the main optomechanical parameters, i.e. the optical mode resonant frequency, the vacuum optomechanical coupling, $\vcr$, the cavity linewidth, $\kappa$ and the cavity outcoupling, $\eta_c$, depend on the position $z_\mathrm{m}$ of the membrane center of mass relative to the intracavity standing wave. 
Since the position of the membrane is constrained by the cavity assembly, we use a laser whose wavelength can be tuned (over $\sim$200nm) to control these parameters. Tuning the laser to the next longitudinal optical resonance introduces one more antinode in the intracavity standing wave, effectively changing the position of the membrane relative to the standing wave.

To predict how the former parameters behave as a function of $\lambda$, we use a transfer matrix model (TMM) \cite{Jayich:2008aa}.  In this approach, the optomechanical system is modeled as a stack of component transfer matrices, whose total behavior can be analyzed to predict system parameters.
We measure the shift of the optical resonant frequencies for several longitudinal modes and fit them with the TMM to estimate a cavity length $L=$~1.6~mm and a membrane position $z_\mathrm{m}=$~0.5~mm relative to the flat, transmissive mirror.
By using an independent measurement of the mirrors' transmissivity, we also predict the cavity outcoupling $\eta_c=\kappa_{\mathrm{out}}/\kappa$ to be modulated between 0.88 and 0.95, depending on membrane position.
With the known laser wavelength, and inferred $z_\mathrm{m}$, the TMM predicts a unique value for $\etac$.
We use this value to obtain an estimate of $\vcr$ from an optomechanically induced transparency (OMIT) trace ($\eta_c$ links measured output power with intracavity photon number), and find excellent agreement with the $\vcr$ obtained from the quantum backaction calibration, which is independent of $\etac$. 
No other results reported in this work depend on $\etac$.

To measure the cavity linewidth $\kappa$ we sweep a phase modulated laser across the cavity resonance and measure the transmitted intensity. The central feature is fitted with a lorentzian and its linewidth is converted to frequency units using the phase modulation sidebands as frequency markers.
During the sweep the auxiliary laser is locked to the red side of a different cavity mode to laser cool the mechanical modes of the entire membrane structure, since these modes' large-amplitude excursions can otherwise lead to artificial broadening of the cavity lineshape.

\subsection{Detection efficiency budget}\label{sec:det_eff_budget}

High detection efficiency is critical for the quantum measurements described in the main text.  
To increase the photodiodes' quantum efficiency, we removed the protective glass window.
From the measured responsivity, we estimate a quantum efficiency of 93\%, which is 2\% below the specified values. 
We believe this difference comes from minor damages during the window removal process. 

\begin{table}
\centering
\begin{tabular}{ l l l }
    \hline
    Optical element & Value & Origin\\
    \hline
    Cavity outcoupling & 95\% & TMM\\
    Cryostat window & 99.6\% & Specs\\
    Lens & 99.6\% & Specs\\
    PBS, transmission & 99\% & Measured\\
    Lens & 99.6\% & Specs\\
    Waveplate & 99.2\% & Specs\\
    PBS, transmission & 99\% & Measured\\
    PBS, reflection & 99.5\% & Specs\\
    Diode quantum efficiency & 93\% & Measured\\
    Interference visibility & 98\% & Measured\\
    \hline\\
\end{tabular}
\caption{\bf Contribution to detection efficiency}
\end{table}
In Supplementary~Table~S2 we report a breakdown of contributions to the total detection efficiency. 
Adding up all losses gives an expected efficiency of $\eta_\mathrm{det}=$~80\%. However, we measure directly the losses between just after the cryostat window and just before the photodetection to be 92\% instead of the expected 95\%, which reduces the detection efficiency to $\eta_\mathrm{det}=$~77\%.
Electronic noise in all measurements is around $1\%$ of the optical vacuum noise level.

\subsection{Calibration of $\vcr$ via quantum noise thermometry or OMIT }\label{sec:OMIT_g0}
We implement two independent methods to measure the single-photon optomechanical coupling $\vcr$ between a given cavity mode and mechanical mode.

\begin{figure}[t]
\begin{center}
\includegraphics[scale=1]{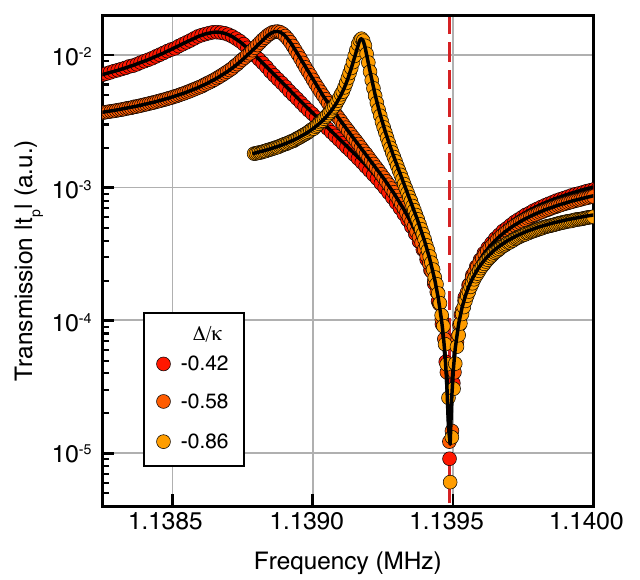}
\caption{{\bf Optomechanical induced transparency (OMIT).}
Measured traces for different laser detuning, close to the mechanical frequency $\Om$ (dashed line). Black lines are theoretical fit.
}
\end{center}
\end{figure}

First, we lock the probe laser on the red side of the cavity and we modulate its phase with a coherent tone. We can observe optomechanically induced transparency\cite{Weis:2010aa} (OMIT) in the optical response function when the frequency of that tone is swept and the intensity of the transmitted beam is directly detected.
If the laser detuning $\Delta$ and the cavity linewidth $\kappa$ are independently measured then the measured trace can be fitted to extract the light-enhanced coupling $g$. From it, the single-photon coupling $\vcr$ can be derived if both the optical losses to the detector and the cavity outcoupling  $\etac$ are known. 

In Supplementary~Fig.~S4 we show a series of OMIT traces with different detuning, with the corresponding fits. While the optical losses in the detection path are measured to be 5\%, the cavity outcoupling cannot be directly measured. Therefore we need to assume the outcoupling coming from TMM for this particular cavity mode, i.e. $\etac=95\%$. With this assumption, we estimate a single-photon coupling of $\vcr/2\pi=129^{+2}_{-3}\,\mathrm{Hz}$.

A second method to measure $\vcr$ is based on the precise knowledge of the temperature of the mechanical mode. When that is the case, one can compare the measured mechanical energy to a known frequency modulation to get $\vcr$. This method is explained more in details in a next section, but we want to point out here that the result from this alternative method is $\vcr/2\pi=(127\pm2)$ Hz. This agreement is particularly meaningful as these methods make very different assumptions.  Put briefly, the first is essentially a calibration based on intracavity {\it photon} occupation, while the second is based on a certain mechanical {\it phonon} occupation.

\section{Theory}
Here we introduce the theoretical model used to analyzed the feedback cooling results.

In Sec.~\ref{sec:feedback_cooling} we describe the dynamics of a mechanical system under the effect of a linear control loop. We give account of the more complex experimental reality and introduce the controller experimentally implemented. We also discuss cooling limits.

In Sec.~\ref{sec:equipart} we show that, in the regime of the experiment, the equipartition theorem is still valid.

\subsection{Feedback cooling}\label{sec:feedback_cooling}

Our model for feedback cooling is a straightforward application of linear control theory, as outlined in the control diagram in Supplementary Fig.~S5.
The mechanical position is estimated from the phase of a resonant probe beam, as described previously. This quantum-optomechanical interaction can be summarized in two main effects in the control diagram. First, this probe necessarily results in a quantum backaction force, $\Fba$, due to radiation pressure shot noise.  Second, optical shot noise sets the fundamental noise floor of the measurement, which can be written in terms of an effective displacement imprecision noise, $x_{\mathrm{imp}}$.  We note that by using a resonant ($\Delta=0$) probe beam, we are able to make the simplifying assumption that $\Fba$ and $x_{\mathrm{imp}}$ are uncorrelated.  In addition to this backaction force, the resonator is driven by a thermal force $\Fth$.  Thus, in the presence of our optomechanical measurement (and in the absence of feedback), the real ($x$) and measured ($y$) mechanical displacement can be written as
\begin{eqnarray}
x(\Omega) &=& \chim(\Og)\Ftot(\Og),\\
y(\Og) &=& x(\Og) + \ximp(\Og),
\end{eqnarray}
where $\Ftot(\Og) = \Fth(\Og) + \Fba(\Og)$ is the total force acting on the system and $\chim(\Og)=[{\meff}(\Om^2-\Og^2-i \Gm\Og)]^{-1}$ is the mechanical susceptibility.

\begin{figure}[hbt]
\begin{center}
\includegraphics[scale=1]{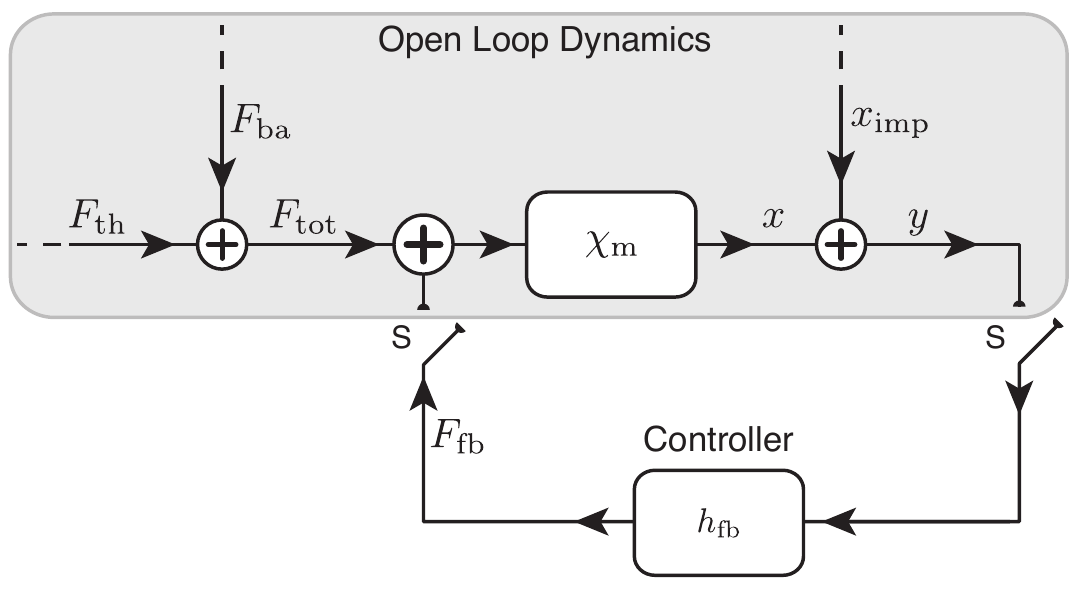}
\caption{{\bf Open and closed loop system dynamics.} Evolution of a mechanical resonator under the action of a thermal reservoir and quantum back-action. The measured position $y$ can be used as an error signal for the controller $\hfb$ to exert an additional force $\Ffb$ on the mechanical resonator, to close the feedback loop and cool the system.}
\label{f:openCloseLoopMR}
\end{center}
\end{figure}

This measurement can also be used as an error signal to tailor an additional, feedback force $\Ffb(\Og)=h_\mathrm{fb}(\Og)y(\Og)$, where $h_\mathrm{fb}(\Og)$ is the controller transfer function.
By closing the loop, i.e. closing the switches S in Supplementary Fig.~S5, this force is exerted on the system and the new dynamics becomes
\begin{eqnarray}
x(\Og) &=& \frac{\chim(\Og)}{1-\chim(\Og)h_\mathrm{fb}(\Og)}\left(\Ftot(\Og)+h_\mathrm{fb}(\Og)\ximp(\Og)\right),\label{eqn:x_closed}\\
y(\Og) &=& \frac{\chim(\Og)\Ftot(\Og)+\ximp(\Og)}{1-\chim(\Og)h_\mathrm{fb}(\Og)}.\label{eqn:y_closed}
\end{eqnarray}
The motional dynamics is modified by the controller in two ways. First, the mechanical response changes and we can introduce the effective susceptibility 
\begin{equation}\label{eqn:chieff_feedback}
  \chieff(\Og)=	\frac{\chim(\Og)}{1-\chim(\Og)h_\mathrm{fb}(\Og)}.
\end{equation}
If the appropriate controller $h_\mathrm{fb}(\Og)$ is chosen, damping and cooling of the mechanical mode can be achieved.
Second, the measurement imprecision noise incoherently drives the system via a force $F_\mathrm{imp}(\Og) = h_\mathrm{fb}(\Og)\ximp(\Og)$, as common to any measurement-based protocol. For large controller gain such that $h_\mathrm{fb}(\Og)\gg\Ftot(\Og)/\ximp(\Og)$, this force starts to heat the mechanical system. 
In the measured displacement $y$ this heating shows up as a squashing of the noise floor, due to correlations between the resonator's motion $x$ and the measurement noise $\ximp$.
These two effects lead to an optimal gain to reach the largest cooling and, at the same time, highlight the importance of having an imprecision $\ximp$ as low as possible.

To quantify these effects, we introduce the double-sided power spectral density of the variable $v(t)$
\begin{equation}
S_{vv}\mathrm{(\Og)}=\int_{-\infty}^{+\infty}\langle v(t)v(0)\rangle e^{\imath\Og t} dt.
\end{equation}
We also define the symmetrized single-sided noise spectrum, $\bar{S}_{vv}\mathrm{(\Og)}=S_{vv}\mathrm{(\Og)} + S_{vv}\mathrm{(-\Og)}$, which represents the entity accessible in the experiment.
The spectrum of the actual and measured displacement, under the loop control, are respectively
\begin{eqnarray}\label{eqn:x_spectrum}
\Sx &=& |\chieff(\Og)|^2\left(\Stot+\left|h_\mathrm{fb}(\Og)\right|^2\Simp\right),\label{eqn:x_spec_closed}\\
\Sy &=& |\chieff(\Og)|^2\left(\Stot+|\chim(\Og)|^{-2}\Simp\right).\label{eqn:y_spec_closed}
\end{eqnarray}
From equation~\eqref{eqn:x_spec_closed} the effective phonon occupancy can be estimated
\begin{eqnarray}\label{eqn:n_from_x_var}
\bn = \frac{1}{2} \left( \frac{\langle\dhp^2\rangle}{2\pzpf^2}+\frac{\langle\dhx^2\rangle}{2\xzpf^2}-1 \right) \approx \int_0^\infty\frac{\Sx}{2\xzpf^2}\frac{d\Og}{2\pi}-\frac{1}{2},
\end{eqnarray}
where the validity of the equipartition theorem is assumed (see Sec.~\ref{sec:equipart} for more details).

\subsubsection{Experimental controller}

In order to reach the quantum regime via feedback cooling, care must be taken in the design of the feedback controller. 
In principle, this involves optimizing the process of state estimation (i.e. how the resonator state is extracted from the measurement record), as well as optimizing the applied force.
Optimal control strategies, both in the classical \cite{Garbini1996aa} and quantum cases \cite{Doherty1999aa,Doherty2000aa,Wiseman2009aa,Hamerly2013aa,Jacobs2014aa,Bowen2015aa} have been developed, and are largely analogous in the LQG regime of (approximately) linear equations of motion, a quadratic cost function, and Gaussian noise.
The controller we employ is inspired by an optimal controller for a single mechanical mode \cite{Garbini1996aa,Doherty1999aa,Jacobs2014aa}, while also addressing the experimental complexity of a multimode system and loop delay. 
In practice, we employ a fourth-order bandpass filter as the main filter, supported by auxiliary filters that address other in-bandgap modes and low-frequency resonances of the membrane.
The total feedback transfer function from the output of the cavity to applied force--including detection electronics, electronic controllers, optical amplitude modulation, and all signal propagation--can be written as
\begin{eqnarray}
\label{eqn:filter_transfer_function}
\hfb(\Og)=\hfbmain(\Og) +\hfbaux(\Og)= \gfb e^{\imath\Og\tau - \imath\phi}\left[\frac{\Gammafb\Og}{\Ogfb^2-\Og^2-\imath\Gammafb\Og}\right]^2+\hfbaux(\Og).
\end{eqnarray}
The center frequency of the main filter is $\Ogfb/2\pi$~=~1.195~MHz and the bandwidth $\Gammafb/2\pi$~=~77.78~kHz.
The overall delay, $\tau$, is measured to be 300 ns. 
The gain, $\gfb$, with unit $\mathrm{kg\cdot Hz^2}$, reflects the electronic gain of the FPGA-based digital filter, as well as the transduction factors associated with the applying a force via amplitude modulation.
In principle, this can be calculated, but we choose to extract this through fits (see Sec.~\ref{sec:feedback_fit}).  
The phase $\phi$ of the filter is adjusted to yield $\arg(\hfb(\Om))\approx -\pi/2$, so that the feedback force on the resonator is then approximately proportional to the latter's velocity, providing a friction force.
Any deviation from $-\pi/2$ will result in a non-zero real part of the transfer function $\hfb$, which, in turn, shifts the resonance frequency of the effective susceptibility (equation~\eqref{eqn:chieff_feedback}). 
The high-order bandpass minimizes the filter gain outside the bandgap, where the system becomes highly multimode.
The onset of feedback-induced instabilities of these out-of-bandgap modes set the practical limit of the gain. 
We also apply one additional narrow filter inside the bandgap, to stabilize a pair of higher-frequency defect modes at $\Og/2\pi\approx$1.25 MHz, and several other filters to stabilize modes outside bandgap, all of which are absorbed into $\hfbaux(\Og)$. 
In Supplementary Fig.~S6, we plot the measured response of the main controller, as well as the main and auxiliary controllers.
The auxiliary filter clearly does not contribute significantly to the controller response near $\Om$.
Note that these measurements only show the controllers' delay of $\sim170\,\mathrm{ns}$, while the total loop delay (including electronic and optical signal propagation etc.) was measured separately to be $\tau=300\,\mathrm{ns}$.
\begin{figure}[hbt]
\centering
\includegraphics[scale=1]{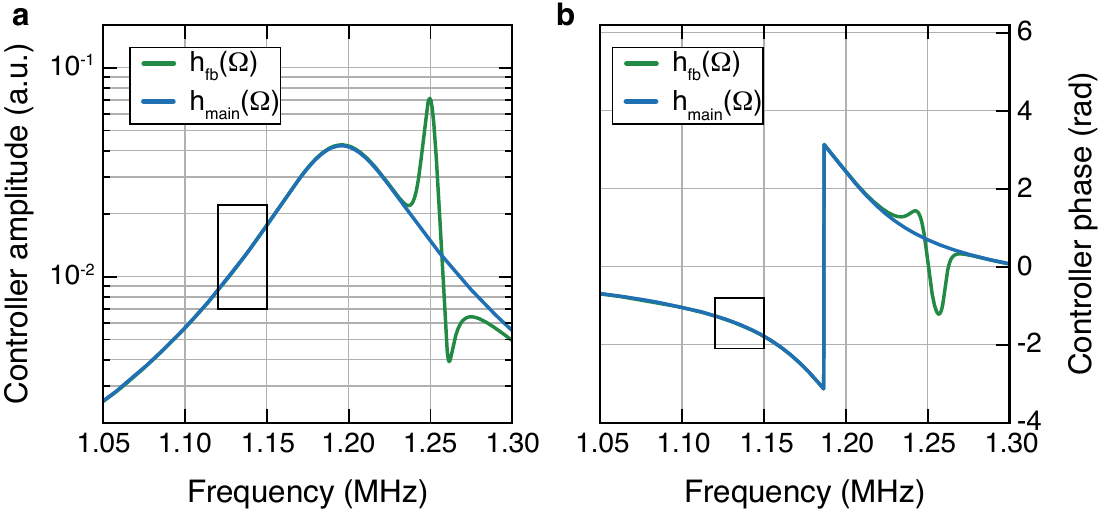}
\caption{{\bf Controller transfer function.} {\bf a, b}, Amplitude and phase response of Red Pitaya.  The blue curve is the response of the main filter only, while the green trace shows also the response of the auxiliary controller. The black rectangle locates the vicinity of the mechanical mode we worked with. The auxiliary controller response is negligible here.}
\label{f:RP_filters}
\end{figure}

\subsubsection{Feedback cooling: occupancy limits}

\begin{figure}[!hbt]
\centering
\includegraphics[scale=1]{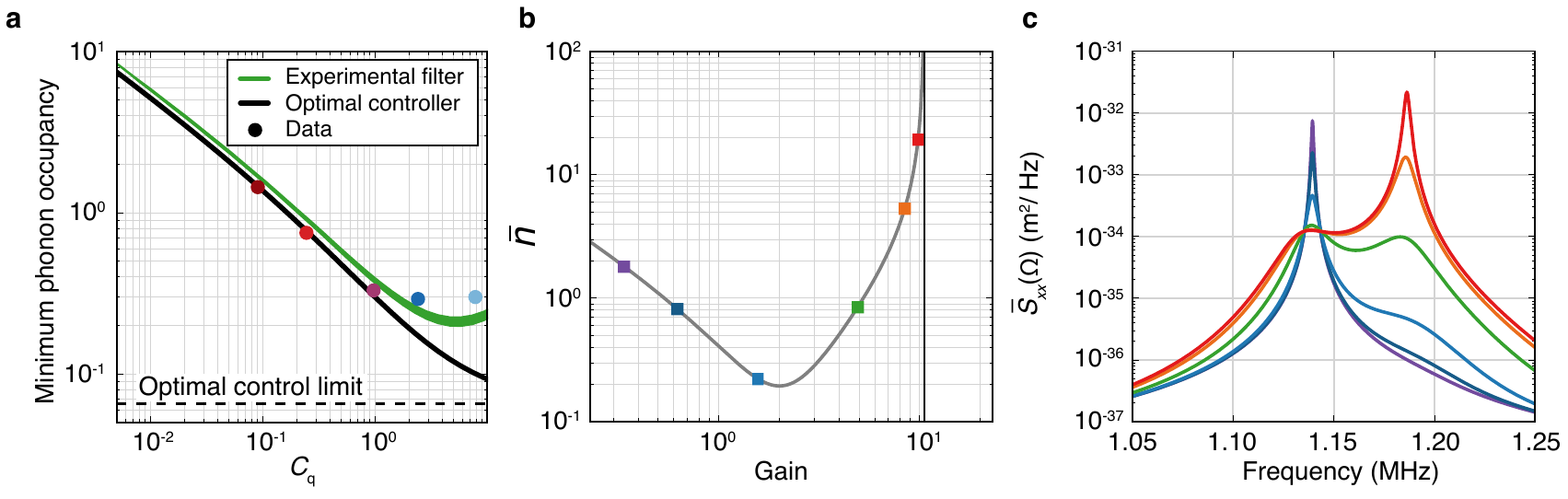}
\caption{{\bf Ultimate feedback cooling limits}. {\bf a}, Minimum occupancy as a function of $\Cq$ for the optimally estimated state (black) and the implemented filter (green). 
The width of the green curve originates in experimental parameter variations. 
Coloured points indicate the measured minimum occupations, as seen in Fig.~4.
The asymptotic limit $\Cq\rightarrow\infty$ of the optimal estimate is marked with a black dashed line.
{\bf b}, Occupancy as a function of gain for $\Cq=10$, showing the onset of an unstable region (hashed area) near $\gfb=10$ 
{\bf c}, Out-of-loop displacement spectra corresponding to the colored markers in {\bf b}, highlighting the eventual instability caused by our filter for very high gains.
}
\label{f:fb_limits}
\end{figure}

The minimum achievable phonon occupancy with feedback cooling depends on the measurement strength ($\Cq$), detection efficiency ($\eta_\mathrm{det}$), and the design of the feedback controller ($h_\mathrm{fb}(\Og)$). 
Ultimately, it is limited, however, by the precision of the estimate of the resonator's motional state \cite{Bowen2015aa,Wiseman2009aa,Jacobs2014aa}.
For our transducer, the optimum estimate's equivalent occupation \cite{Bowen2015aa} can be approximated to (assuming a single-mode, high-Q resonator coupled to a hot thermal bath)
\begin{equation}
\label{eq:fb_limit}
 \bnest \approx \frac{1}{2}\sqrt{\frac{\Cq+1}{\eta_\mathrm{det} \Cq}}-\frac{1}{2} = \frac{1}{2}\left(\sqrt{\frac{1}{\eta}}-1\right).
\end{equation}

Our experimentally-implemented bandpass filter provides an approximation of the optimum filter, but does show a deviation for large $\Cq$. 
Supplementary Figure~S7a shows the prediction of equation~\eqref{eq:fb_limit}, compared with the numerically-calculated limits of the filter we implement. Supplementary Fig.~S7b and c also illustrate the source of this deviation.
In Supplementary Fig.~S7b, we see the phonon occupation as a function of gain for $\Cq$=10.
Coloured markers at several gain values correspond to out-of-loop displacement spectra $\Sx$ shown in Supplementary Fig.~S7c.
We see that at sufficiently high gain the filter leads to an instability near its center frequency. 

\subsection{Equipartition condition}\label{sec:equipart}

It has been shown that application of feedback forces can result in deviations from the equipartition theorem, particularly if the bandwidth of the feedback is not constrained (as in the ``cold-damping'')\cite{Genes2008aa, Habibi2016aa}.
This means that a na{\"i}ve calculation of the position variance {only, as in equation~\eqref{eqn:n_from_x_var},} can underestimate the energy of the resonator.
In our measurement, the use of a high-order bandpass filter ensures that there is no unbounded divergence of the momentum variance, as predicted in the ``cold-damping'' model.
To confirm that there is no finite-frequency discrepancy between position/momentum variance, we explicitly calculate the momentum spectral density in terms of our previously derived position spectral density $\Sp=m \Og^2 \Sx$. 
Supplementary Fig.~S8 shows $\Sp$ and $\Sx$ for the cooperativity and gain corresponding to the minimum observed occupation reported in the main text. By integration of these spectra, we find that the variances in position and momentum differ by less than $1\% $.

\begin{figure}[h]
\begin{center}
\includegraphics[scale=1]{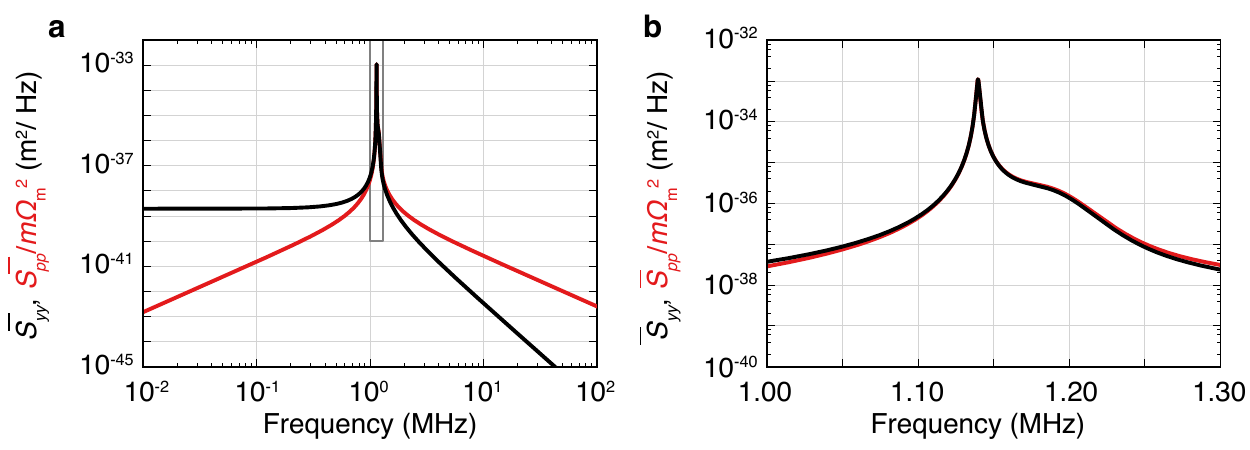}
\caption{{\bf Consistency with equipartition.} {\bf a}, Wide and {\bf b}, narrow  views of $\Sp$ and $\Sx$ for $\Cq=$~2.4 and the optimum gain (i.e. the settings which produced the minimum occupation from the main text.}
\label{f:equipartition}
\end{center}
\end{figure}

\clearpage
\newpage
\section{Data Analysis}
Here we describe in more details the data analysis performed to obtained the results shown in the main text.

In Sec.~\ref{sec:g0_calib} we report how the sideband cooling experiment to the quantum regime has been analyzed and the procedure to extract $\vcr$ from it.

In Sec.~\ref{sec:effect_detuning} we analyze how a finite, non-zero detuning affects the position measurements.

In Sec.~\ref{sec:feedback_fit} we give a description of the fit of the in-loop spectra of the feedback cooling and the consequent estimation of the effective phonon occupancy.

Finally, in Sec.~\ref{sec:si_heating_rate} we describe how the heating rate is extracted from a measurement of a transient dynamics, due to thermal heating.

\subsection{Quantum backaction limit and calibration of $\vcr$}\label{sec:g0_calib}
Here we give an overview of the analysis of the sideband cooling experiment in Fig.~2.
The mechanical motion is monitored via the phase fluctuations of a weak ($\Cq\approx$~0.08), resonant probe laser ($\kpr/2\pi=$~15.9~MHz) while the strong auxiliary laser is locked on the red side ($\Daux/2\pi=$~-4.2~MHz) of another cavity mode ($\kaux/2\pi=$~12.9~MHz) for sideband cooling (and imparting quantum backaction, as well will see). 
PSDs of the recorded voltage fluctuations, for increasing power of the auxiliary beam ($\Paux$), are shown in Supplementary Fig.~S9a, zoomed in around $\Om$. The auxiliary beam results in broadening of the peak (optical damping), as well as a shift in mechanical frequency (optical spring)\cite{{Aspelmeyer2014aa}}. 
We fit each spectrum with a Lorentzian model to retrieve the effective linewidth $\Geff$ and resonance frequency shift $\delta\Om$. They show a linear trend as a function of cooling power $\Paux$ (Supplementary Fig.~S9b and c) as predicted by a standard model for optomechanical dynamical backaction:
\begin{eqnarray}\label{eqn:freq_shift}
\delta\Om &=& \Oeff - \Om = \mathrm{g_{aux}}^2\left[\frac{\Daux+\Om}{(\Daux + \Om)^2 + (\kaux/2)^2}+\frac{\Daux-\Om}{(\Daux - \Om)^2 + (\kaux/2)^2}\right],\\
\Geff &=& \Gm+\Gopt = \Gm + \mathrm{g_{aux}}^2\left[\frac{\kaux}{(\Daux + \Om)^2 + (\kaux/2)^2}-\frac{\kaux}{(\Daux - \Om)^2 + (\kaux/2)^2}\right].\nonumber\\\label{eqn:opt_damping}
\end{eqnarray}
$\Paux$ refers to the power transmitted through the cavity, such that the light-enhanced optomechanical coupling strength is given by $\mathrm{g^2_{aux}}=\vcrAux^2\Paux/\left(\hbar\Omega_\mathrm{L}\kaux\eta_\mathrm{c,aux}\right)$.
The data in Supplementary Fig.~S9b and c are fit with equations~\eqref{eqn:freq_shift} and \eqref{eqn:opt_damping}, respectively. From these fits, we extract a light-enhanced coupling of $\mathrm{g_{aux}}/2\pi \approx$~24~kHz  for a cooling power of $\Paux$~=~1~$\mathrm{\mu W}$.
\begin{figure}[ht!]
\begin{center}
\includegraphics[scale=1]{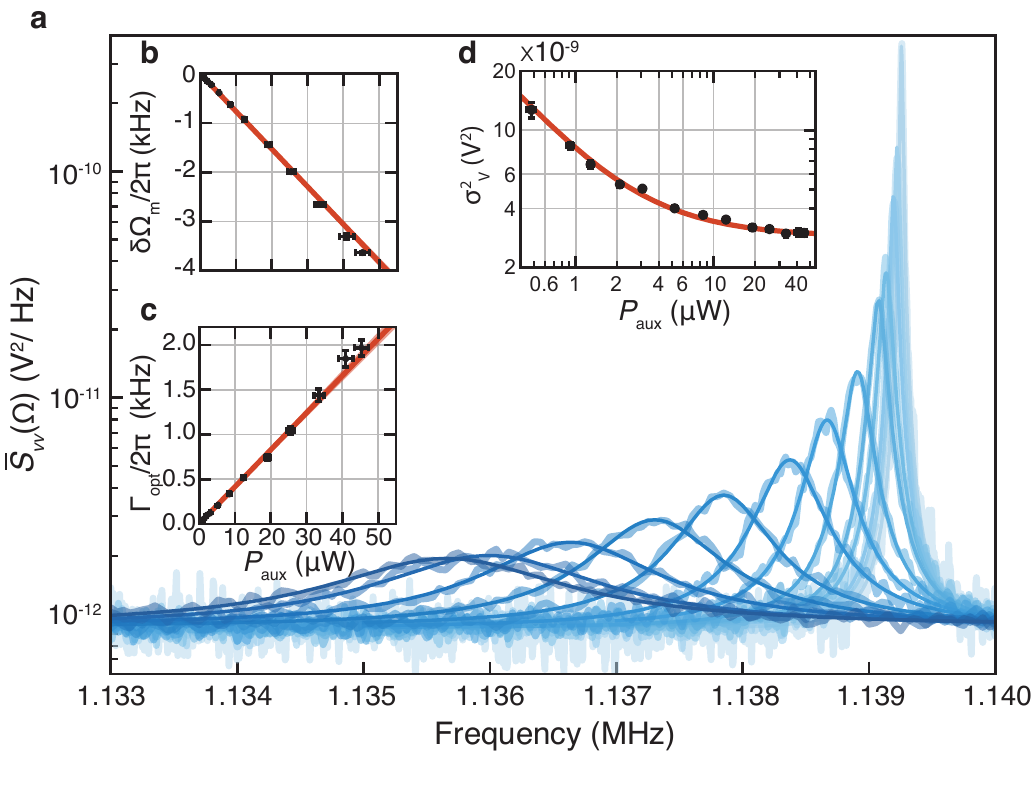}
\caption{{\bf Calibration of $\mathrm{g_{aux}}$ and $\vcr$.} {\bf a}, Voltage noise spectra of the probe beam phase fluctuations for different cooling power $\Paux$, increasing from light to dark blue. The solid lines are lorentzian fits, from which we estimate the effective resonance frequency $\Oeff$ and linewidth $\Geff$.
{\bf b, c}, Resonance shift  $\delta\Og$ and $\Geff$, respectively, as a function of $\Paux$. The solid red lines are fits based on the dynamical backaction model, while their width reflects the statistical uncertainties of the fit parameters.
{\bf d}, Area under the mechanical peak, from Lorentzian fits.}
\label{f:g0calibration}
\end{center}
\end{figure}

Initially, this optical damping results in a cooling of the effective temperature of the mechanical resonator. 
However, a full quantum treatment reveals that backaction of the laser provides an additional source of force noise.  
At some point, these cooling and heating processes equilibrate, resulting in a minimum achievable phonon occupancy (the ``sideband cooling limit''). 
These processes are captured by the following expression for the effective phonon occupancy\cite{Aspelmeyer2014aa}:
\begin{equation}
\bar{n} = \frac{\Gm\bnth + \Gopt\bnmin}{\Gm+\Gopt},
\end{equation}
where $\bnmin = \left((\kaux/2)^2+(\Daux+\Om)^2\right)/\left(-4\Daux\Om\right)$ is the sideband cooling limit.
Note that this assumes that any other source of heating is negligible, e.g. laser classical noise (see Sec.~\ref{sec:class_noise}) and quantum backaction from the probe laser. 

We make use of the fact that we can deeply access the quantum backaction regime, in which $\bn\rightarrow\bnmin$, for a robust calibration of $\vcr$ {\it of the probe laser} used to monitor the mechanical fluctuations induced by the auxiliary laser's quantum backaction.
In particular, for the largest auxiliary laser power, we have $(1+\Gopt \bnmin/\Gm \bnth)^{-1}\sim{4\%}$.
That is, down to a small correction, the phonon occupation is determined by the optical parameters $\kaux$, $\Daux$ and $\Om$ only, which can be easily and robustly determined spectroscopically.
We can then use a frequency modulation technique\cite{Gorodetksy2010aa} to translate the measured homodyne voltage fluctuations into cavity frequency fluctuations, which must obey $\langle \delta \oC^2 \rangle=\vcr^2 (2 \bn+1)$.

In practice, we integrate the voltage noise spectrum around $\Om$ to get the variance $\mathrm{\sigma^2_V}$. 
This can be related to the phonon occupancy by $\mathrm{\sigma^2_v} = (\mathrm{K}(\Om)/\Om^2)\vcr^2 \left(2\bar{n} + 1\right)$.
The transduction function $\mathrm{K}(\Om)$, between the measured voltage noise and the cavity frequency fluctuations, can be obtained by frequency-modulating the laser with known depth, and measuring the induced voltage modulation in the homodyne receiver\cite{Gorodetksy2010aa}.
To obtain the most accurate value for $\vcr$, we take the data points with all $\Gopt$, obtained at different auxiliary laser powers, and fit them to the model 
\begin{eqnarray}\label{eqn:link_n_area}
\mathrm{\sigma^2_v}(\Paux) = 2\frac{\mathrm{K}(\Om)}{\Om^2} \vcr^2 \left(\frac{\Gm\bnth + \Gopt(\Paux)\bnmin}{\Gm + \Gopt(\Paux)} + \frac{1}{2}\right)
\end{eqnarray}
with the only unknown parameters $\bnth$ and $\vcrP$, which are, then, used as fit parameters.
The other parameters $\Om$, $\kaux$, $\Daux$ are directly measured, $\mathrm{K}(\Om)$ obtained by frequency modulation calibration \cite{Gorodetksy2010aa} and, for the lower auxiliary powers, $\mathrm{g_{aux}}$ which comes from the fits in Supplementary Fig.~S9b and c.
Since we can access both the classical and the quantum regime of sideband cooling, we can extract both  $\Tm=(11\pm2)\,\mathrm{K}$ and $\vcrP/2\pi=(127\pm2)\,\mathrm{Hz}$ from the fit.

The value obtained for  $\vcr$ in this way is very well compatible with $\vcr/2\pi={129^{+2}_{-3}\,\mathrm{Hz}}$ determined from an OMIT fit, which requires knowledge of the intracavity photon number, and therefore $\etac$--but does not depend, e.g., on laser noise.
In all analyses of phonon occupancies (imprecision, backaction, or steady-state mechanical occupancy) we use the same frequency modulation calibration technique \cite{Gorodetksy2010aa} to calibrate the raw spectra.
Using this together with the $\vcrP/2\pi=(127\pm2)\,\mathrm{Hz}$ obtained in the same calibration implies that in the finally stated phonon occupancies, the calibration factor $\mathrm{K}(\Og)$ cancels.
In other words, all occupancies are eventually referenced to the equilibrium occupancy obtained in the quantum backaction limit $\bnmin$.
Spectra shown in the main manuscript are re-expressed in units of $\mathrm{m^2/Hz}$ mainly to facilitate interpretation.
For this we use a $\xzpf$ with the simulated mass as stated in Supplementary Table~S1.

\subsection{Effect of finite detuning}\label{sec:effect_detuning}

\begin{figure}[ht!]
\centering
\includegraphics[scale=1]{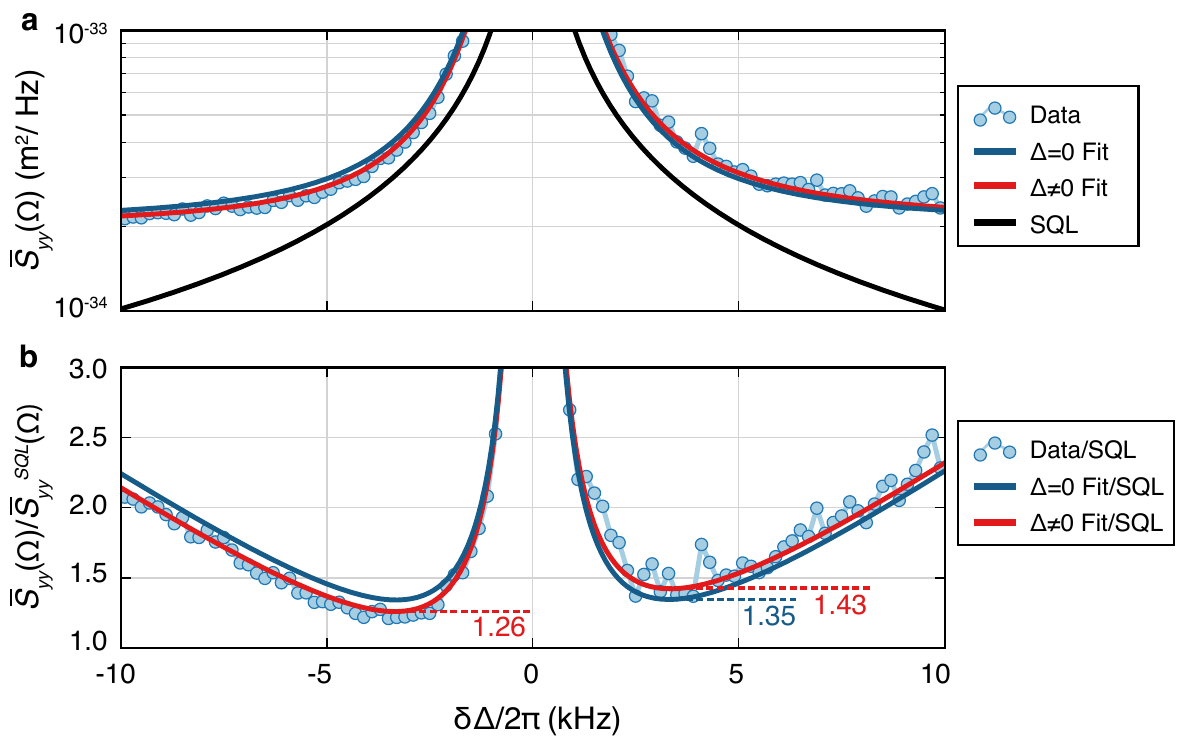}		
\caption{{\bf Effect of finite detuning.} {\bf a}, The displacement spectra corresponding to the closest approach to the SQL (off the mechanical resonance), as reported in the main text. Data is shown, along fits as described in the text.  The frequency-dependent SQL is shown for reference.
{\bf b}, The same data and fits, divided by the SQL.  The minimum values of both the $\Delta=0$ and $\Delta\neq 0$ fits are indicated.}
\label{f:finiteDetuning}
\end{figure}

In the main text, we describe a position measurement based on the phase quadrature of a resonant ($\Delta=0$) optical probe.  In such a measurement, the quantum fluctuations which drive the mechanical motion are uncorrelated with those forming the imprecision noise floor (as the effects originate in amplitude and phase fluctuations, respectively).  This is no longer true whenever the optical quadratures are mixed, either by cavity rotation (when $\Delta\neq0$) or by measurement of a different quadrature at the homodyne detector. Recent works have highlighted the ability of such correlations to improve displacement detection\cite{Kampel:2017aa,Sudhir:2017aa}, particularly for frequencies away from the mechanical resonance, as in Fig.~3.  In principle, such correlations can even allow sub-SQL sensitivity, though this has not yet been achieved experimentally.

While such effects are not prominent in our measurements, a careful analysis reveals that a small non-zero detuning of our probe beam (previously estimated to be $\Delta\approx-0.05\kappa$) does result in some correlation between the motion and noise floor. Supplementary Fig.~S10a shows the highest-$\Cq$ displacement spectrum (i.e. the darkest blue curve from Fig.~3a and b), along with the SQL spectrum, $\bar S_{yy}^\mathrm{SQL}(\Omega)=2\hbar |\chi_m(\Omega)|$. We note that this SQL curve is the lowest possible, corresponding to a perfect phase measurement of a resonant probe.  Note also that the calibration of the displacement spectra (in absolute units) is still done via the same phase calibration technique, which is still valid for non-zero detuning.  The dark blue line is a fit to a simple Lorentzian curve (i.e the expected lineshape for $\Delta=0$).  We note that the data possess a slight asymmetry below/above the mechanical resonance frequency.  To properly account for this, we can instead describe the data by a simple model including cavity-induced quadrature rotation\cite{Nielsen2017aa}, allowing the probe detuning to take on some small, non-zero value.  The red curve is the result of fitting to such a model.  In the $\Delta=0$ analysis, we used the effective bath occupation ($n_{\mathrm{tot}}$) and noise floor ($n_{\mathrm{imp}}$) as fit parameters in a generic Lorentzian lineshape (equation~\eqref{eqn:y_spec_closed} with $\hfb=0$).  Here, the cavity-induced rotation model takes only fundamental system parameters as input.  Thus, in analogy with the Lorentzian fit, we allow the probe strength and detection efficiency to vary as fit parameters.  The fit results correspond to total force noise and imprecision noise which match independent predictions to within factors of 1.08 and 1.07, respectively.  The probe detuning is also used as a fit parameter, yielding $\Delta=-0.04\kappa$, consistent with other estimates based on probe transmission and dynamical backaction.

Supplementary Fig.~S6b shows these same data and fits, relative to the SQL spectrum.  The minima of such a curve corresponds to the best performance of the measurement, in terms of sensitivity near the SQL. The minima of the $\Delta=0$ fit is 1.35, as reported in the main text.  The $\Delta\neq0$ model indicates an improved performance below the mechanical resonance, and a degraded performance above, reaching minima of 1.26 and 1.43, respectively.  Thus, for measurement below the mechanical resonance, we find a displacement sensitivity which is in fact only 26$\%$ higher than the SQL.  For clarity in the main text, we report the simplified $\Delta=0$ result, which also corresponds to the average of the below- and above-resonance figures-of-merit from the full $\Delta\neq0$ model.

\subsection{Feedback cooling fitting}\label{sec:feedback_fit}
In Fig.~4 we have presented the results of feedback cooling measurements at various values of $\Cq$. Supplementary Fig.~S11 illustrates the analysis procedure/results for one such $\Cq$ (the one producing the lowest phonon occupancy, i.e. $\Cq=$~2.4).  The process consists of fitting the in-loop displacement spectrum to equation~\eqref{eqn:y_spec_closed} for $\Sy$, then inferring the out-of-loop spectrum $\Sx$ and integrating to obtain the phonon occupancy.  The filter gain ($g$) and phase ($\phi$) are used as fit parameters, as well as $\nimp$ and $n_\mathrm{tot}$.  These fit results are presented in Supplementary Fig.~S11a, as a function of the digital gain set on the Red Pitaya.  As expected, $g$ is proportional to the digital gain and all spectra are consistent with a constant phase.  Moreover, all spectra are consistent with a constant value of $\nimp$ and $n_{\mathrm{tot}}$, indicating that no gain-dependent electronic noise is added by our feedback loop. 
Supplementary Fig.~S11b shows the measured spectra at various levels of feedback gain, highlighting that the spectra are well-described by our model. The resonance peak shifts towards higher frequency as the filter gain $\gfb$ increases. This is due to a negative, non-zero, real part of the transfer function $\hfb$, as measured in Supplementary Fig.~S6.
Supplementary Fig.~S11c shows examples of in- and out-of-loop spectra for low (red) and high (blue) values of the gain.  Note that at low gain, the in-loop spectra simply adds a noise floor to the out-of-loop spectra.  At high gain, correlations between the motion and noise floor result in destructive interference, or squashing.

\begin{figure}[hbt]
\includegraphics[scale=1]{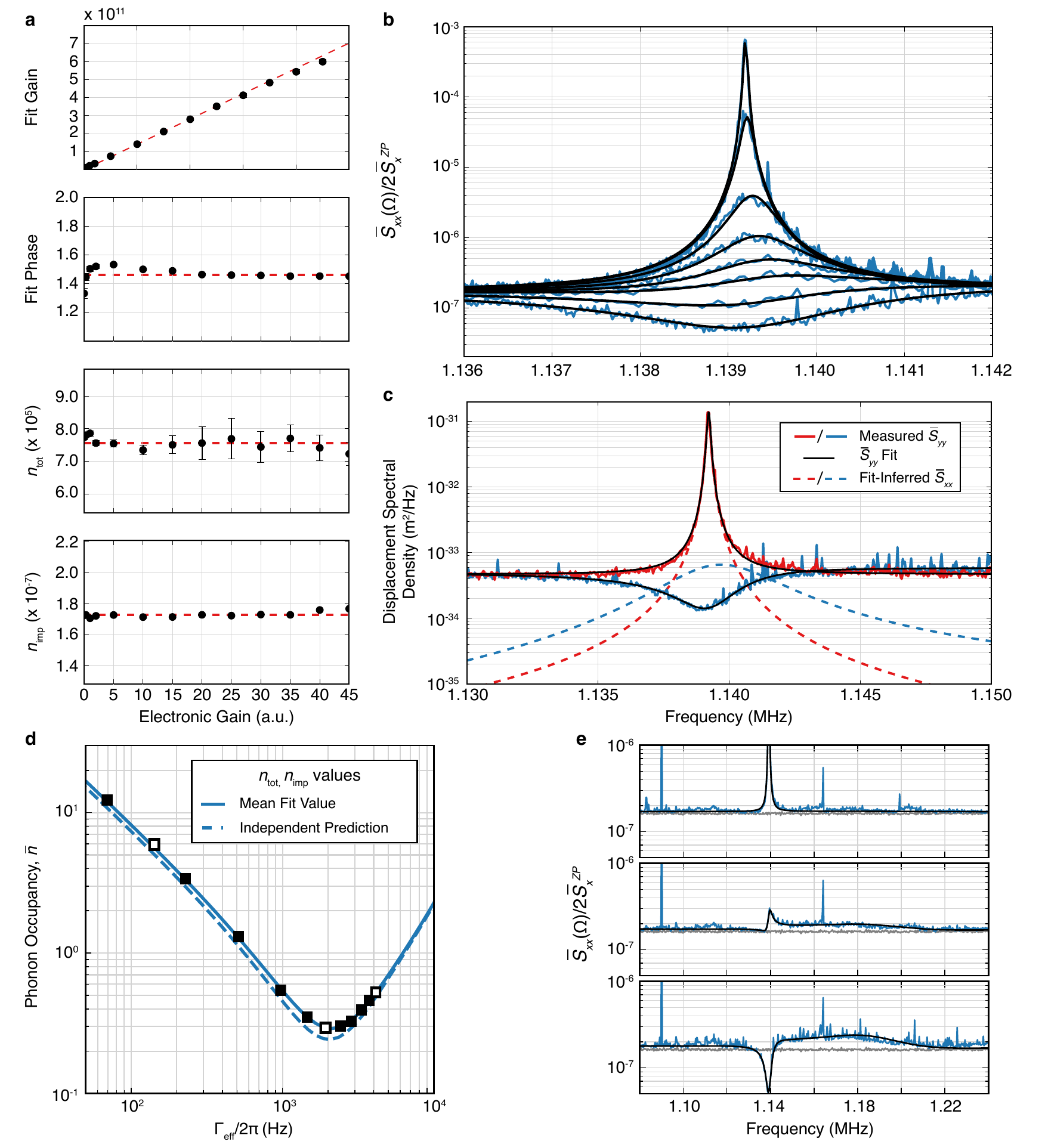}
\caption{{\bf Sample feedback cooling analysis for $\mathbf{\Cq=}$~2.4.} {\bf a}, Fit results, with constant/linear fits shown as red lines.
{\bf b}, Fits of in-loop spectra for electronic gains between 0 and 45.
{\bf c}, In-loop spectra and fits, with calculated out-of-loop spectra for low (red) and high (blue) gain.
{\bf d}, Comparison of extracted phonon occupancies with theoretical curves based on independent parameter predictions and fit values (i.e. the red lines in {\bf a}).
{\bf e}, In-loop spectra for three values of gain (as marked in {\bf d}), shown over a wider range, highlighting the ability of our model to describe excess motion caused by our filter at high gains.
}
\label{f:feedbackAnalysis}
\end{figure}
\clearpage

\subsection{Heating rate}\label{sec:si_heating_rate}
Figure~4 also reports a measurement of the mechanical heating rate out of the ground state.
We apply a feedback scheme (see Sec.~\ref{sec:feedback_cooling}) to cool the resonator to a low phonon occupancy state, i.e. $\nI\approx2$.
By means of an electronic switch in the $h_\mathrm{main}(\Og)$ controller path, the feedback force is turned off, thus leaving the resonator free to evolve towards a new (warmer) thermal equilibrium\cite{Pinard2000aa,Kleckner2006aa}, with $\nF\approx60$, due to residual sideband cooling.
This process is done repeatedly, so that the experimental results can be averaged. 
The repetition rate is set by the 200~ms-period square wave which drives the switch.
We note that this switch only closes the feedback controller cooling the defect mode of interest, whereas all the other channels $h_\mathrm{aux}(\Og)$ are always on to stabilize other modes of the entire $\mathrm{Si_3N_4}$ structure.
The homodyne photocurrent is sampled both from a DAQ, providing a wide-bandwidth FFT, and from a digital lockin-amplifier (LIA), providing a signal which is demodulated at $\Om$, and filtered with a bandwidth of 300~Hz.

To analyze this non-stationary heating process, we record the time-dependence of the mechanical motion using the LIA during a full switching period in terms of the measured quadratures $\Xt$ and $\Yt$.
Then, we average over 400 experimental runs to get the time dependent variance $\RftSqAv = \XftSqAv + \YftSqAv$.
This variance $\RftSqAv$ still contains detection shot noise. To remove this, we independently measure the (stationary) shot noise variance $\RsnftSqAv$ by blocking the signal arm to the homodyne detector.
Ultimately, the heating process is described by $\finRftSqAv = \RftSqAv - \RsnftSqAv$, shown in Supplementary Fig.~S12a.

\begin{figure}[ht]
\begin{center}
\includegraphics[width=0.75\linewidth]{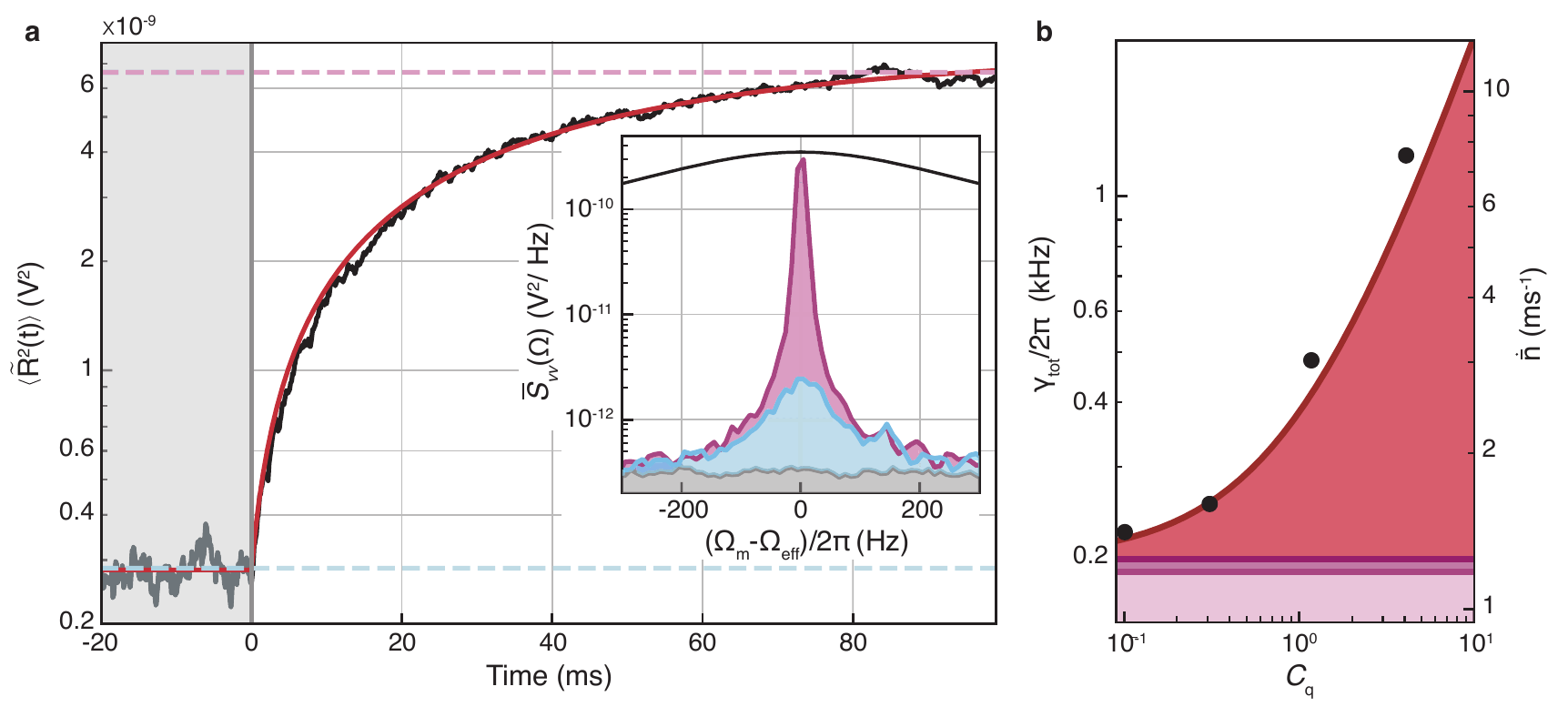}
\caption{{\bf Mechanical state evolution due to thermal bath heating.} {\bf a}, Measured evolution (red trace) of the  averaged variance.
The grey region indicates that the feedback force is on. 
The light blue (purple) horizontal dashed lines is an estimation of $\nI$ ($\nF$), coming from numerical integration of the spectra (see inset) acquired from the DAQ, with the feedback controller permanently on (off).
The black line in the inset represents the filter gain response employed by the lock-in amplifier.
{\bf b}, Measured total heating rate $\gtot$ at different probe beam $\Cq$ (black symbol). The solid red line is an independent prediction.
Three main contributions manifest: the thermal decoherence (light purple area), the optical decoherence from the from the probe laser (red area) and a small decoherence from the auxiliary laser (dark purple).
} 
\label{f:heating_rate_si}
\end{center}
\end{figure}

To correctly interpret these data, a careful calibration has to be carried out.
First, we convert $\finRftSqAv$ from a voltage noise $[\mathrm{V^2}]$ to a phonon occupancy ${\overline{n}}$ based on equation~\eqref{eqn:link_n_area}, just as we did in Sec.~\ref{sec:g0_calib}. This relies on $\vcrP$ (which we already know, from Sec. \ref{sec:g0_calib}) and $\mathrm{K(\Om)}$, which can be obtained from a frequency modulation calibration technique done in the spectral domain\cite{Gorodetksy2010aa}. To measure it, we turn off the feedback controller $h_\mathrm{main}(\Og)$ and compute the photocurrent PSD, shown as a dark blue trace in the inset of Supplementary Fig.~S12a (the calibration tone, at $\Og/2\pi$~=~1.09~MHz, is not shown in the plot).
Second, since this conversion factor comes from data acquired with the DAQ, while $\finRftSqAv$ comes from the LIA, we need to assess any electronic imperfection which can lead to different, absolute level of the voltage noise as measured by the LIA compared to the DAQ if we want to apply this conversion factor to $\finRftSqAv$. To correct the signal coming out from the LIA, we use as a marker the optical shot noise, measured by blocking the signal arm. Then, the DAQ spectrum is filtered in the same way as the LIA data, and the integrated variance is used to calibrate the LIA voltage.

Once the measured process is calibrated as described, we fit it to a heating model
\begin{equation}
\overline{n}(t) = \nI + \theta(t)(\nF-\nI)\left(1-e^{-\Geff\,t}\right),
\end{equation}
where $\theta(t)$ is the Heaviside function, $\nI$ and  $\nF$ are the, respectively, initial and final phonon occupancy and $\Geff$ is the mechanical linewidth in the absence of feedback. The time origin has been defined such that at $t=$~0~ms the feedback is switched off.
From the fitted parameters we can extrapolate a coherence time (inverse heating rate out of the ground state) $1/\gtot=1/(\Geff\,\nF)\approx$~730~$\mathrm{\mu s}$.
A more direct way to estimate the coherence time is by measuring the initial slope of the heating process. In fact $\dot{\bar{n}}(t)|_{t=0} = (\nF-\nI)\Geff\approx\gamma_\mathrm{tot}$ if $\nF\gg\nI$.

The continuous presence of the probe and auxiliary lasers introduces additional decoherence, degrading the expected thermal coherence time $1/\gamma\approx$~850~$\mathrm{\mu s}$. 
This is shown in  Supplementary Fig.~S12b, where the measurement described is repeated for different $\Cq$ of the probe beam. The behaviour of the data can be well explained by a prediction, which takes into account the contribution to the decoherence both of the fixed auxiliary beam and of the probe beam.
\clearpage
\section{Classical Laser Noise}\label{sec:class_noise}
Here we show and describe the measurements and techniques adopted to characterize the classical noise of the laser used in the experiment.
To ensure quantum-limited measurements, it is important to characterize the classical amplitude and phase noise of the laser running the experiment.

To measure the amplitude noise, we directly shine the laser onto a photodiode and record the photocurrent PSD, at different optical power levels.
To extract the classical contribution at a given Fourier frequency $\Og$, we integrate the PSDs in a bandwidth of 20~kHz around $\Omega$ and plot the resulting variance as a function of the optical power impinging on the detector, as shown by the red symbols in Supplementary Fig.~S13b. 
We fit the data to a parabola with the linear term coming from the shot noise of the light, while the quadratic term corresponds to classical noise.
The ratio of the quadratic term to the linear one indicates the contribution of the classical noise in units of shot noise per optical power, at a given $\Og$. By repeating this procedure at different $\Og$, we obtain the curve shown in in Supplementary Fig.~S9a. The vertical axis is expressed relative to the shot noise of 1~$\mathrm{\mu W}$, e.g. for 1~$\mathrm{\mu W}$ of impinging light power the classical noise contribution is, at $\Omega/2\pi = $~1.14~MHz, 0.08\% of the corresponding shot noise level.

To measure the phase noise, we send the laser light through a low-finesse optical cavity with a linewidth of $\kappa/2\pi = $~2.44~MHz.
When the laser is detuned from the cavity, the input phase noise is rotated into amplitude noise by the cavity. We detect these amplitude fluctuations by direct direction of the transmitted light with a photodiode. To extract the classical phase noise contribution, we record the photocurrent PSD at different detuning $\Delta$ and, as for the amplitude noise, we integrate them in a bandwidth of 20~kHz around $\Omega$. In Supplementary Fig.~S13d we plot the resulting variances (red symbols) as a function of detuning $\Delta$, with a fit line coming from a simple model\cite{Galatola:1991aa,Zhang:1995aa}, given by: 
\begin{equation}\label{eqn:clas_noise_spec}
\bar{S}_{XX}^\mathrm{out} = 1+\frac{4(1-\eta_c) \eta_c  \kappa ^2}{\Delta ^2+(\kappa/2) ^2}\frac{\left((\Delta^2 + (\kappa/2)^2)^2+ (\kappa/2) ^2 \Omega ^2\right)\Cx+\Delta ^2 \Omega^2\Cy}{\Delta ^4+2 \Delta ^2
   \left((\kappa/2) ^2- \Omega ^2\right)+\left((\kappa/2) ^2+ \Omega ^2\right)^2},
\end{equation}
where $\Cx$ and $\Cy$ are, respectively, the classical amplitude and phase noise normalized to the shot noise level, which is 1. In the fit we fix the amplitude noise to the previously measured value, thus leaving the phase noise $\Cy$ as the only free parameter. By repeating this procedure at different $\Og$, we obtain the curve shown in Supplementary Fig.~S13c, with the vertical axis units expressed, again, relative to the shot noise of 1~$\mathrm{\mu W}$.
With these measurements of the laser amplitude and phase noise, it is possible to calculate\cite{Jayich:2012aa} their relative contribution to the heating of the oscillator.  In the measurement of Fig.~2, we estimate that classical amplitude (phase) noise of the auxiliary beam contributes less than $<3\%$ ($<1\%$) to the total decoherence at the maximum power.  In the strong probe measurement experiment of Fig.~3, we estimate both amplitude and phase contributions to the total decoherence to be less than 1$\%$.  For this measurement, we also estimate that phase and amplitude noise contributions to the measurement noise floor should both be significantly less than 1$\%$.
\begin{figure}[ht!]
\begin{center}
\includegraphics[scale=1]{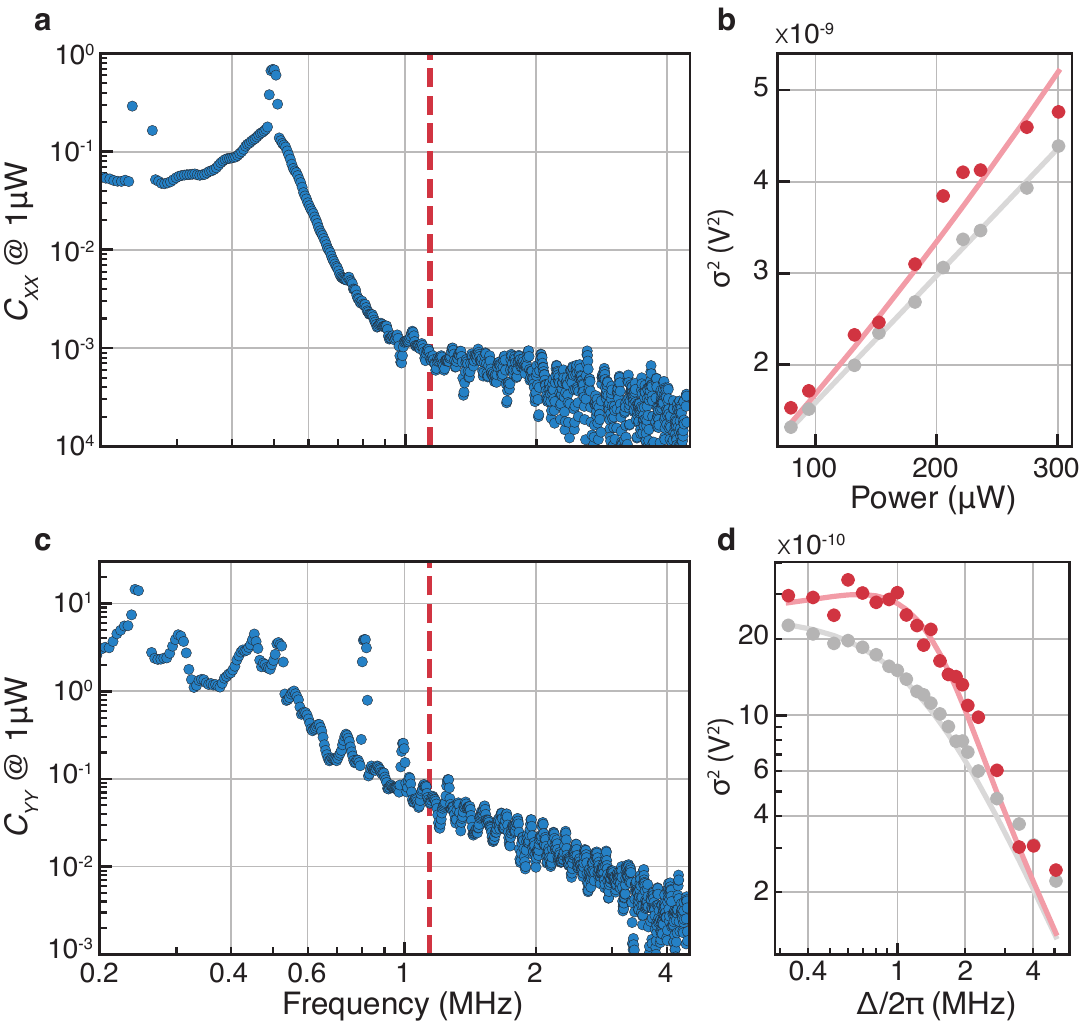}
\caption{{\bf Laser classical amplitude and phase noise.} {\bf a}, Amplitude noise as a function of Fourier frequency. The dashed red line marks the mechanical resonance frequency $\Om$. {\bf b}, Variance at $\Om/2\pi=$~1.14~MHz (red symbols) as a function optical power. The solid red line is a quadratic fit. The gray symbols come from integrating the PSD at 4.5~MHz, where the classical noise is negligible (i.e. they are dominated by shot noise.) The solid gray line is a linear fit. {\bf c}, Phase noise as a function of Fourier frequency. The dashed red line marks $\Om$. {\bf d}, Variance at $\Om$ (red symbols) as a function detuning $\Delta$. The solid red line is a fit from a cavity rotation model. The gray symbols come from integrating the PSD at 4.5~MHz, where the classical noise is negligible, i.e. they are dominated by shot noise. The solid gray line is a fit from the same model but without classical noise.}
\label{f:class_noise}
\end{center}
\end{figure}

\end{document}